\documentclass[a4paper,12pt]{article}

\usepackage{amssymb,amsmath,mathbbol,mathrsfs}
\usepackage[usenames,dvipsnames]{color}
\usepackage{hyperref}
\usepackage{xcolor}
\usepackage{stmaryrd}
\usepackage{authblk}
\usepackage{framed}
\usepackage{empheq}
\usepackage{slashed}
\usepackage{hyphenat}
\usepackage{cite}
\usepackage{scalerel}[2016/12/29]
\usepackage{marginnote}    
\usepackage[left=.8in,right=.8in,top=.79in,bottom=.79in]{geometry}                  
\hypersetup{
	colorlinks=true,         
	linkcolor=MidnightBlue,          
	citecolor=BrickRed,        
	urlcolor=MidnightBlue            
}
\usepackage{sectsty}
\allsectionsfont{\bfseries\sffamily} 

\newcommand{\BIBand}{and}

\usepackage{cite}

\usepackage{amssymb,amsmath,amsthm,mathbbol,mathrsfs,mathtools,slashed,mathrsfs}
\usepackage{stmaryrd}
\usepackage{xcolor}
\usepackage{enumerate}
\usepackage{marginnote}
\usepackage{longtable}

\newcommand{\be}{\begin{equation}}
\newcommand{\ee}{\end{equation}}

\newcommand{\Lie}{\mathrm{Lie}}

\newcommand{\A}{{\mathcal{A}}}
\renewcommand{\d}{{\mathsf{d}}}
\newcommand{\D}{{\mathsf{D}}}

\newcommand{\SU}{{\mathrm{SU}}}

\newcommand{\G}{{\mathcal{G}}}
\newcommand{\Ad}{{\mathsf{Ad}}}
\newcommand{\pp}{{\partial}}

\renewcommand{\bar}{\overline}
\newcommand{\dd}{{\mathbb{d}}}

\renewcommand{\hat}{\widehat}

\newcommand{\T}{{\mathsf{T}}}

\newcommand{\SO}{\mathrm{SO}}
\renewcommand{\ker}{{\mathrm{ker}}}
\newcommand{\slashf}{//_{\hspace{-3pt} f\hspace{2pt}}}

\newcommand{\rad}{{\text{rad}}}
\newcommand{\Coul}{{\text{Coul}}}
\newcommand{\GC}{{\mathsf{G}}}

\newcommand{\cint}{{\int\kern-.87em{<}}}
\newcommand{\sint}{{\int\kern-.75em{\sim}}}
\newcommand{\fint}{{\int\kern-1.00em{\int}}}

\newcommand{\bb}{\mathbb}

\newcommand{\tr}{\mathrm{Tr}}
\renewcommand{\div}{\mathrm{div}}

\renewcommand{\i}{(\textit{i}) }
\newcommand{\ii}{(\textit{ii}) }
\newcommand{\iii}{(\textit{iii}) }

\newcounter{Ex}[section]
\renewcommand{\theEx}{\arabic{section}.\arabic{Ex}}
\newenvironment{Ex}[1][]{\refstepcounter{Ex}\par\medskip
   \noindent \textbf{Example~\theEx} (\textit{#1})  \rmfamily }{\medskip\hfill$\bullet$}

\newcounter{Rmk}[section]
\renewcommand{\theEx}{\arabic{section}.\arabic{Ex}}
\newenvironment{Rmk}[1][]{\refstepcounter{Ex}\par\medskip
   \noindent \textbf{Remark~\theEx} (\textit{#1})  \rmfamily }{\medskip\hfill$\bullet$}

\newcounter{Aeq}

\newcounter{Caux}

\renewcommand{\#}{\sharp}

\let\oldmarginpar\marginpar
\renewcommand\marginpar[1]{\oldmarginpar{\color{red}\raggedright\footnotesize #1}}

\usepackage{enumitem}

\title{\sffamily Edge modes without edge modes}
\author[1]{\sffamily Aldo Riello\thanks{Email: aldo.riello [at] ulb [dot] be}}
\affil[1]{\small  Physique Théorique et Mathématique, Université libre de Bruxelles, Campus Plaine C.P. 231, B-1050 Bruxelles, Belgium}

\begin{document}
\maketitle

\abstract{
We discuss gauge theories of the Yang--Mills kind in finite regions with boundaries, and in particular the definition of the corresponding quasi-local degrees of freedom and their gluing upon composition of the underlying regions.
Although the most of the technical results presented here has appeared in previous works by Gomes, Hopfmüller and the author, we adopt here a new perspective.
Focusing on Maxwell theory as our model theory, in most of the text we avoid technical complications and focus on the conceptual issues related to symplectic reduction in finite and bounded regions, and to gluing---e.g. superselection sectors, non-locality, Dirac's dressing of charged fields, and edge modes. 
In this regard, the title refers to a gluing formula for the reduced symplectic structures, where the ``edge mode'' contribution is explicitly computed in terms of gauge-invariant bulk variables. 
Despite capturing most interesting features, the Abelian theory misses some crucial technical and conceptual points which are present in the non-Abelian case. To fill this gap, we dedicate the last section to a brief overview of functional connection forms, flux rotations, and geometric BRST, among other topics.

%
 }


{\hypersetup{	linkcolor=black, }\tableofcontents}
%
\vfill
{\hypersetup{	linkcolor=MidnightBlue }}



\section{Introduction} 

Note. \emph{The material presented in this article heavily draws in its technical aspects on previous work by the author with Gomes and also Hopfmüller, to which this article still refers for the proof of certain statements (among the works cited below, see in particular \cite{GomesHopfRiello, GomesRiello-quasilocal, AldoNew}). The organization and exposition of the material are novel, and so are many remarks.}\\

In the last five years, interest in gauge theories on manifolds with boundaries has surged. 
This has been largely, although not solely, a consequence of new insights into the symmetry structure of Yang-Mills and gravitational theories at asymptotic infinity (see \cite{strominger2018lectures} and references therein for an overview).
Efforts to understand these new structures---together with a general interest in understanding holography and entanglement entropy---have led many authors to investigate the interplay of gauge and diffeomorphism symmetries not only with asymptotic boundaries (like in the classic works of \cite{ReggeTeitelboim1974, WaldZoupas, BarnichBrandt:2001, Carlip1995}) but also with finite ones \cite{Donnelly:2014fua, AronWill, DonnellyFreidel, Delcamp:2016eya, Speranza:2017gxd, Geiller:2019bti, Chandrasekaran:2019ewn, Harlow_cov, Mnev:2019ejh, DonFreidSper:2020, RielloSoft, GomesRiello2016, GomesRiello2018, GomesHopfRiello, GomesRiello-quasilocal, AldoNew} (to mention just a few recent ones).\footnote{Two other lines of research in the mathematical foundations of quantum field theory (QFT) with boundaries (of all co-dimensions) that must be mentioned are: the algebraic extended-TQFT framework for topological theories \cite{Baez:1995xq,Lurie:2009keu}, and the BV-BFV approach to perturbative (and \emph{non}-necessarily topological) QFT \cite{cattaneo2014classical, Cattaneo:2015vsa}. }

In this context, a large part of the discussion is framed in terms of so-called ``edge modes,'' which one can (loosely) characterize as extra boundary degrees of freedom (dof) introduced to resolve the conflict between gauge and boundaries \cite{Balachandran:1994up, Carlip1995, DonnellyFreidel}. 
In this context, a lot of the intuition comes from the bulk-boundary relationship between the Chern--Simons and Wess--Zumino--Witten theories\footnote{This bulk-boundary relationship is physically realized in the effective description of the quantum Hall effect \cite{FrohlichZee:1991}.} \cite{Witten:Jones, Gawedzki:1999bq, Balachandran:1994up, Carlip1995}, especially as formulated in \cite{Balachandran:1994up}.
This intuition is often, if somewhat implicitly, used to justify the introduction of edge dof in \emph{any} gauge theory, and irrespectively of whether the boundaries are physical or fiducial (by ``fiducial'' we mean boundaries that are ``drawn in the air'' and not physically identified with the edge of a medium characterized by its own specific internal dof and boundary conditions).
This generic conclusion might raise some questions in particular in the case of Yang--Mills theory which, contrary to Chern--Simons, $BF$, or gravitational theories, has a gauge-invariant Lagrangian \emph{density} \cite{GomesHopfRiello, GriffinSchiavina}.

The goal of this article is to provide a succinct yet rigorous account of the dof content in Yang--Mills theory over (finite and) bounded regions. 
To avoid confusions, it is important to stress the following:
\begin{enumerate}[label=\roman*.,font=\itshape]
\item our analysis focuses on Yang--Mills theories and is not directly transposed to Chern--Simons, $BF$, or gravitational theories;
\item the focus is on the Yang--Mills dof supported in a (finite and) bounded region $R\subset \Sigma$, where $\Sigma$ is a Cauchy surface.  
\end{enumerate}
The latter point means that we are ultimately concerned with the dof associated to the causal domain (or causal ``diamond'') $D(R)$---and \emph{not} with those associated to the spacetime cylinder $C\cong R\times\bb R$ which best abstracts the spacetime evolution of a system ``in a box.''
In this sense, our setup is best understood in terms of fiducial, rather than physical, boundaries.\footnote{A third natural possibility exists: studying fields over $D(R)$ but with a focus on the evolution along the null (future) boundary $\pp D(R)^+$. See e.g. \cite{Wieland:2020gno} where this approach is applied to asymptotic general relativity.}

There are two crucial differences between the $D(R)$ and $C$ set-ups. 
First, whereas $D(R)$ is uniquely determined as a spacetime region by $R\subset\Sigma$, the spacetime cylinder $C$ is not, it requires some notion of the evolution of the boundary, i.e. of the time-evolution of the ``box'' the system is thought to live in.
Second, and relatedly, whereas $D(R)$ supports an autonomous dynamics for the associated dof which is uniquely specified by a set of initial conditions at $R$, in order to fully fix the dynamics over $C$ one has to carefully specify a set of boundary conditions (and fluxes) at the boundary $\pp C \cong \pp R \times \bb R$ \cite{Harlow_cov} (e.g. as in a Casimir setup \cite{BarnichCasimir:2019}). For example, when studying entanglement entropy, the relevant notion of subsystem is the one associated to $D(R)$.
({Extra} boundary conditions can always be further imposed at the ``belt'' $\pp R$ of the causal domain $D(R)$, but these simply restrict the phase space of interest, and are not instrumental to its very definition.)

In this article, we will completely characterize the gauge-invariant dof of Yang--Mills theories over $D(R)$. 
We will do so in a non-manifestly spacetime-covariant fashion, i.e. in a canonical setting.
More specifically, we will develop a theory of canonical symplectic reduction for the Yang--Mills degrees of freedom over $R\subset \Sigma$.

In studying symplectic reduction in the presence of boundaries, we will insist on a third, fundamental, point:
\begin{enumerate}[label=\roman*.,font=\itshape]\setcounter{enumi}{2}
\item gauge at $\pp R$ will be treated on the same footing as gauge in the bulk of $R$, that is as a mere redundancy in the description of the system.
\label{iii}
\end{enumerate}
In other words, we will adopt the time-honoured perspective that gauge is a mere redundancy in the description of a gauge theory's dof; a redundancy which must be introduced to describe in a local fashion a set of dof which is otherwise intrinsically \emph{nonlocal} \cite{DeWitt:1962mg}. Crucially, in Yang--Mills theory, we will find that this approach is perfectly consistent and does not ``miss'' any dof; more on this below.\footnote{A subtle caveat to this statement is provided by topology-supported dof; more on this below.}

Nonlocality is intrinsic to any theory subject to a \emph{Gauss constraint}. 
Indeed, the Gauss constraint is truly the main character on the stage of Yang--Mills theories \cite[Ch.7]{StrocchiBook} (and \cite{strocchi2015symmetries} for a non-technical discussion), even when its role is seemingly overshadowed by a focus on gauge freedom and invariance.
The viewpoint adopted in this work de-emphasizes the role played by gauge freedom in favour of the crucial role played by the Gauss constraint---\emph{especially} in relation to boundaries. 

Our focus on the Gauss constraint and its implications in the presence of boundaries is a defining feature of the approach, which distinguishes it from most of the recent literature. Moreover, our focus on the Gauss constraint is the reason why large swaths of  our analysis are tailored to gauge theories of the Yang--Mills kind, and not directly exportable to, say, Chern--Simons theories where the functional property of the gauge-generating constraint (flatness) are quite different.

The most important consequence of our attentive analysis of the Gauss constraint in the presence of boundaries---supported by other evidence that we will discuss in due time---is the existence of super-selection sectors associated to specific values of the electric flux $f$ through the boundary $\pp R$.\footnote{See the Outlook section of \cite{GomesRiello-quasilocal} for a brief comparison of the quasilocal superselection of the electric flux discussed here, and the one arising at asymptotic (spatial) infinity first discussed in the algebraic QFT literature.}
That is, in a bounded region the electric flux fails to correspond to a (dynamical) dof and the theory naturally ``factorizes'' into sectors of prescribed electric flux. 

Importantly, the superselection of $f$ is \emph{not} a statement about some boundary dof, but rather a statement about the (symplectic) properties of the Coulombic electric field \emph{throughout} $R$.
This is most clearly seen in pure Maxwell theory (i.e. in Abelian charge-less Yang--Mills theory), where---\emph{on-shell of the Gauss constraint}---$f$ entirely determines the Coulombic electric field throughout $R$, via
$$
E_\text{Coul} = \pp \varphi
\qquad\text{and}\qquad
\begin{cases}
\Delta \varphi = 0 & \text{in }R,\\
\pp_\perp \varphi = f & \text{at }\pp R.
\end{cases}
$$
Therefore, if $f$ must be treated as an external parameter for the theory in $D(R)$, so must $E_\text{Coul}=\pp \varphi$.

(Although  for simplicity we define here the Coulombic component of the electric field as a pure-gradient, i.e. in ``Coulomb gauge,'' we will later discuss not only how analogous statements hold in any other gauge, but also in which sense all these statements have entirely gauge-invariant consequences when it comes to the definition of the reduced symplectic structure. Cf. Remark \ref{Rmk:Erad} and Section \ref{sec:YM}.)

These statements and conclusions cannot be transposed verbatim to the non-Abelian case: there the electric flux fails to be a gauge-invariant quantity and therefore cannot be ``fixed'' without breaking gauge invariance. The solution to this issue is not completely straightforward and requires us to \textit{i.} partition the space of on-shell field configurations (on-shell of Gauss, that is) into spaces characterized by a given electric flux specified \emph{up-to-gauge}---we call these spaces \emph{covariant superselection sectors}---and \textit{ii.}  introduce a ``{completion}'' of  the naive symplectic structure over each covariant superselection sector, since the latter happens to be degenerate. In a sense, the degeneracy of the naive symplectic structure is due to the fact that the flux is not completely fixed in a covariant superselection sector, and is therefore a characteristic of the non-Abelian theory only.
 
Our symplectic ``completion'' satisfies two crucial properties.
First, it is fully canonical: it does not depend on any arbitrary choice and is instead given to us once a covariant superselection sector is fixed (i.e. once a gauge-conjugacy class of electric fluxes is chosen).
And second, it does not entail the enlargement of the underlying phase space of on-shell Yang--Mills configurations modulo gauge, but simply prescribes a way to define a viable symplectic structure on that very space.

We emphasize this second point because it underscores an attitude which is somewhat orthogonal to the ``edge mode'' approach, where the original phase space \emph{is} enlarged by the introduction of extra boundary dof; we will come back to edge modes shortly. 

The fact that in a covariant superselection sector the non-Abelian fluxes are free to vary within a given conjugacy class, brings about new features that do not arise in the Abelian setup. In particular, it is now possible to ``rotate'' the fluxes within their conjugacy class, hence altering the entire Coulombic electric field, while keeping the rest of the field content, e.g. the radiative and matter dof, completely fixed. Since these transformations do not rotate \emph{all} fields at the same time, they cannot be pure-gauge. They are, in fact, physical and descend onto the reduced phase space. 

The most obvious consequence of these transformations, which we call ``flux rotations,'' is the resulting Poisson non-commutativity of the electric fluxes in a given covariant superselection sector  \cite{CattaneoPerez} (this has a well-known analogue on the lattice). 

Moreover, we will see that the flux rotations come in different flavours, depending on how they are extended into the bulk (via a prescription to define the Coulombic electric field throughout $R$), and only certain flavours correspond to Hamiltonian transformations in the reduced phase space. Interestingly, this property of theirs depends on the global properties of the corresponding extensions \emph{within phase-space}.

\begin{center}
$\ast\quad\ast\quad\ast$
\end{center}

So far we have discussed the properties of the Yang--Mills dof  within a given region $R$, and argued that it naturally leads to the phenomenon of flux superselection.
Superselection can be summarized as the fact that the electric flux $f$ through $\pp R$, and hence the entire Coulombic electric field within $R$, is a non-dynamical dof (with respect to the region $D(R)$).

The fact that certain (physical!) components of the electric field drop from the (dynamical) phase space raises the question of whether or not the combination of the (dynamical) gauge-invariant dof supported in $R$ and $\bar R = \Sigma\setminus R$ comprises the entirety of the gauge-invariant dof supported in $\Sigma$.
As common, we name the problem of reconstructing the entirety of the dof over $\Sigma$ from those supported on $R$ and $\bar R$, the \emph{gluing problem}. 

(Note that, for simplicity, we neglect here the possibility of topology-supported dof, such as Aharonov--Bohm phases, i.e. nontrivial holonomies and monodromies which are supported by non-contractible cycles of $\Sigma$. Degrees of freedom of this kind, especially when not contained in either $R$ nor $\bar R$, are clearly not ``reconstructible'' from the regional dof supported in $R$ and $\bar R$. For more on the emergence of Aharonov-Bohm dof upon gluing within the formalism proposed in this article, see \cite[Sect.6.8]{GomesRiello-quasilocal}.)

At this point, however, we find valuable to further split the gluing problem into two related but distinct questions.

The \emph{first} gluing problem asks:
\begin{enumerate}[label=\arabic*.,font=\itshape]
\item  is it possible to reconstruct all the gauge-invariant dof supported over $\Sigma$ as function(al)s of the gauge-invariant (dynamical) dof supporrted on $R$ and $\bar R$---in spite of the superselection of the electric flux? 
\end{enumerate}
Maybe surprisingly, the answer to the first gluing problem is `\emph{yes},' and indeed we will provide explicit formulas that achive this (highly nonlocal) reconstruction.
Hence, superselection does not spoil gluing.

Then, in light of the this reconstruction, the \emph{second} gluing problem asks:
\begin{enumerate}[label=\arabic*.,font=\itshape]
\setcounter{enumi}{1}
\item  does the reduced symplectic structure over $\Sigma$ factorize into the sum of the reduced symplectic structures over $R$ and $\bar R$? 
\end{enumerate}
The answer to the second gluing problem is `\emph{no},' and we will explicitly identify the new term in the symplectic structure that prevents the factorizability. 

%

This extra term does \emph{not} feature an extra pair of coupled dof. Instead, it introduces \emph{a nonlocal symplectic coupling between the reduced gauge-invariant dof already present in the two complementary regions}. 

The structure of this coupling is reminiscent of that featured in the edge mode phase space of \cite{DonnellyFreidel}, where the electric flux is the conjugate variable to a would-be-gauge dof. 
Indeed, the symplectic-coupling term involves the electric flux through the interface $S = \pm\pp R^\pm$ (once again understood as a proxy for the Coulombic electric field over $R$ and/or $\bar R$) together with another term that superficially looks like a pure-gauge phase. However, the crucial point here is that both the flux and this ``phase'' do \emph{not} constitute a new independent pair of dof, but---by the answer to the first gluing problem---they are \emph{function(al)s} of the gauge-invariant (dynamical) dof supported on {both} $R$ and $\bar R$. 
The form taken by these functionals will be made explicit in the main text.

Therefore, the non-factorizability of the global symplectic structure is not due to some ``missing'' interface dof, but to a nonlocal symplectic coupling between the gauge-invariant dof associated to each region.

This fundamental difference  not only explains the meaning of this article's title, \emph{Edge Modes Without Edge Modes} (see Section \ref{sec:EMWEM} for more details), but it also sheds further light on the superselection of the electric flux.

Since the \emph{reconstructed} electric flux is a highly nonlocal functional of the the gauge-invariant (dynamical) dof supported on \emph{both} $R$ and $\bar R$, it cannot be determined in terms of the gauge-invariant (dynamical) dof supported solely on $R$, say.
Therefore, when focusing on the intrinsic dof supported on $R$ (and evolving within $D(R)$), the flux $f$ must be externally prescribed, as a minimal proxy for the information we are losing by ignoring the dof supported over $\bar R$ (dof which matter because of the nonlocality introduced by the Gauss constraint). 
Compellingly, this viewpoint matches all the crucial features of the computation of entanglement entropy in lattice gauge-theory \cite{DonnellyEntEnt:2011}: not only is the superselection of the electric flux\footnote{On the lattice, it is possible to give a magnetic superselection prescription too (at least in 2+1d or in the Abelian theory \cite{Casini_gauge,radicevic2014notes,Delcamp:2016eya}); however, this symmetry is broken in our treatment once we commit to describing the phase space of Yang--Mills theory in terms of gauge potentials.} through the entangling surface revealed upon tracing out the dof supported over one potion of the lattice, but also the weight associated to each (covariant) superselection sector is reflected in our prescription for the symplectic-srtucture's ``completion'' discussed above.\footnote{This second statement won't be discussed in detail later, so let us sketch the argument here. On the lattice, each electric flux up-to-gauge corresponds to a choice of irrep of $G$, the charge group of the gauge theory under investigation. The entanglement entropy associated with a given superselection sector $f$ then contains a non-distillable contribution measuring the total ``size'' of the super-selection sector, defined in terms of the dimensions of the relevant flux irreps (see the second term on the rhs of equation (32) of \cite{DonnellyEntEnt:2011}). In turn,  via Kirillov's coadjoint orbit method, the dimension of this Hilbert space matches precisely the symplectic volume associated to a covariant superselection sector via (a descretized version of) the KKS completion of the symplectic structure mentioned above, cf. \eqref{eq:KKScomplete}. For more on this, see \cite[Sec.7]{GomesRiello-quasilocal}.}

It is instructive to briefly contrast this state of affairs with gluing for a scalar field theory: there the dof are purely local, and the answer to both gluing problems is (somewhat trivially) `yes.' In Yang--Mills theory, it is the nonlocal nature of the coupling between the dof supported on $R$ and $\bar R$, which eventually leads to the superselection of the flux and to the nonfactorizability of the symplectic structure with respect to the decomposition $\Sigma = R \cup \bar R$. This nonlocality is in turn a direct consequence of the Gauss constraint. 

The picture painted here supports the relational perspective on gauge theories \cite{RovelliGauge2013, Gomes:2019otw},\footnote{The article \cite{RovelliGauge2013} first laid out the conceptual basis of the relational approach to gauge theories, of which our formalism can be seen as an explicit instantiation. However, there---and especially in a recent follow-up \cite{Rovelli:2020mpk}---certain statements are made that we do not agree with, in particular about what is measurable in a gauge theory. For the relevance of our formalism to relationalism, see Section \ref{sec:Gluing} on gluing. For yet another viewpoint on these matters, see \cite{Teh2015}.} and bears important analogies with the theory of dynamical (quantum) reference frames \cite{Bartlett:2007zz, Vanrietvelde:2018pgb, Hoehn:2019owq, Hoehn:2021wet} (see \cite{Aharonov1967SSR,aharonov1967:obs,Bartlett:2007zz} for more on superselection in this context). In fact, the nonlocality inherent in a gauge theory implies, via dynamical and symplectic coupling, that the dof over one regions serve as ``reference frames'' for the dof over its complement. It is tempting to speculate that these analogies might turn into something more concrete when the ideas developed here for Maxwell and Yang--Mills theories are applied to general relativity \cite{GomesRiello2016}.

\begin{center}
$\ast\quad\ast\quad\ast$
\end{center}

To conclude this introduction, one word on edge-modes.
In the language of symplectic reductiton, edge modes à la \cite{DonnellyFreidel} arise if one decides to reduce by bulk-supported gauge-transformations \emph{only}, that is only by gauge transformations which are trivial at the boundary---in ``violation'' of our point \textit{\ref{iii}} above.
Reduction by a smaller gauge group results in a reduced phase space which is larger and contains extra dof with respect to the one discussed so far.
One important and subtle point, which is often not recognized, is that the edge-mode construction \emph{breaks} gauge-invariance at the boundary: the resulting symplectic form depends on some arbitrary choice (implicitly) made at $\pp R$.
We refer to \cite[Sec.5]{AldoNew} for a more comprehensive discussion of this point. 

\paragraph*{Organization of the article}

This article is organized as follows. 
In order to rigorously discuss the notions of ``phase space'' and of ``degree of freedom,'' we need to use the language of symplectic geometry.
And to rigorously deal with the notion of ``phase space of gauge-invariant degree of freedom,'' we need to use the formalism of symplectic reduction. For the article to be self contained, and fix notations as well, we give a brief review of these standard concepts in Section \ref{sec:thingssymp} and \ref{sec:MaxNoBdry}, respectively. 

In section \ref{sec:thingssymp}, we also review the definition of a symplectic foliation, which is necessary to give a mathematical account of the notion of superselection sector.

Sections \ref{sec:MaxBdry} and \ref{sec:Gluing} are dedicated to symplectic reduction in the presence of boundaries, and to gluing respectively. Both these sections will focus on pure Maxwell theory---that is on an Abelian Yang--Mills theory with no charged matter.

In Section \ref{sec:MaxMatt}, we will briefly discuss the main features of reduction and gluing in the presence of matter, but still in the Abelian theory. One of the features we will discuss is the role of (Dirac) dressing of charged matter fields within the formalism. Another one is the emergence of a subtle, and physically relevant, ambiguity in gluing associated with the total electric charge of a set of particles contained in $R$ ($\bar R$ must contain the opposite amount of charge).

Finally, in Section \ref{sec:YM} we will discuss how the non-Abelian theory differs from the Abelian one both from a physical and mathematical perspective. Among other things, we will give a brief overview of the technical tools---in particular the field-space connection form---needed for addressing the reduction and gluing in the non-Abelian context. 

We conclude the article with a comment on the functional connection and geometric BRST. 


\paragraph*{Acknowledgements}
My gratitude goes to P. Höhn for his kind invitation to present this work at OIST, Japan, albeit remotely due to the COVID-19 pandemic, not least because the preparation of that series of talks directly led me to the write these notes. I would also like to thank H. Gomes for his feedback on this article, and for the many conversations we had over the years on the topics treated here.
This project has received funding from the European Union’s Horizon 2020 research and innovation programme under the Marie Sk{\l}odowska-Curie grant agreement No 801505.

\section{A flash review of things symplectic}\label{sec:thingssymp}

This section provides a flash overview of the concepts in symplectic geometry that will be needed in the rest of the article.

\paragraph{Phase spaces and symplectic geometry}

The physical concept of a ``phase space'' corresponds to the mathematical concept of a \textit{symplectic manifold} \cite{Arnold_1978}. 
A manifold $\Phi$ is said symplectic if it equipped with a 2-form $\Omega$,\footnote{Throughout this article I will use Einstein's summation convention over repeated indices, unless otherwise explicitly stated.}
\be
\Omega = \Omega_{IJ} \d z^I \wedge \d z^J \in \Omega^2(\Phi),
\ee
which is
\begin{enumerate}[label=\roman*.,font=\itshape]
\item non-degenerate, $\ker (\Omega_{IJ}) = 0$, 
\item and closed, $\d \Omega = 0$.
\end{enumerate}
The relation between the symplectic 2-form $\Omega$ and the maybe more common Poisson bracket on $\Phi$ is readily provided once the Poisson bracket is expressed in terms of a bivector $\Pi \in \mathfrak{X}^{\wedge 2}(\Phi)$:
\be
\{\cdot\,,\cdot\} = \Pi^{IJ} \frac{\pp}{\pp z^I}\otimes\frac{\pp}{\pp z^J}, \qquad \Pi^{IJ} = (\Omega_{IJ})^{-1}.
\label{eq:PoissonBracket}
\ee
The bivector $\Pi$ is then \i non-degenerate, \ii antisymmetric, and \iii satisfies the following identity (Jacobi):
\be
\Pi^{IL}\frac{\pp}{\pp z^L}\Pi^{JK} +  \Pi^{JL}\frac{\pp}{\pp z^L}\Pi^{KI} +  \Pi^{KL}\frac{\pp}{\pp z^L}\Pi^{IJ} 	= 0.
\ee
One can show that this identity follows from the closedness of $\Omega$.

\begin{Ex}[Cotangent bundle]
Given a manifold $Q$, its cotangent bundle $\Phi = \T^*Q$ is canonically symplectic. Indeed, $\T^*Q$ carries the \textit{canonical 1-form} $\theta$, which can be written in adapted coordinates
 $z^I = (q^i, p_i)$ as:\footnote{The canonical, or tautological, 1-form is the unique 1-form $\theta \in \Omega^1(\T^*Q)$ which ``cancels pullback." That is, it is the unique 1-tform on $\T^*Q$ such that, for any $\alpha\in\Omega^1(Q)\simeq \Gamma(\T^*Q\to Q)$, $\alpha =\alpha^*\theta$. In adapted coordinates $(q^i, p_i)$, the 1-form $\alpha = \alpha_i \d q^i$ defines the section $\alpha: Q\to \T^*Q$, $q^i \mapsto (q^i, p_i = \alpha_i)$, so that $\alpha = \alpha_i \d q^i = \alpha^*\theta $.}
\be
\theta =  p_i \d q^i.
\ee
The differential of this 1-form, gives then the \textit{canonical 2-form} $\Omega$,
\be
\Omega = \d \theta = \d p_i \wedge \d q^i,
\label{eq:DarbouxOm}
\ee
which is manifestly symplectic. In the same coordinates, the corresponding Poisson bracket, is the canonical Poisson bracket
\be
\{\cdot\,,\cdot\} = \frac{\pp}{\pp q^i}\otimes\frac{\pp}{\pp p_i} - \frac{\pp}{\pp p_i}\otimes\frac{\pp}{\pp q^i} .
\label{eq:DarbouxPi}
\ee
\end{Ex}

A manifold $\Phi$ is said a \emph{Poisson manifold} if it is equipped with a bivector $\Pi$ which satisfies condition \ii and \iii above \cite{FernandesBookPoisson}. In other words,  Poisson manifolds generalize symplectic manifolds by allowing for possibly degenerate Poisson bivectors.

\paragraph{Hamiltonian symmetries}
On a symplectic manifold $(\Phi,\Omega)$, to each function $f: \Phi \to \bb R$ corresponds a vector field $X_f$:
\be
X_f = \{ f, \cdot \}
\qquad\text{or equivalently}\qquad
\mathsf{i}_{X_f} \Omega = - \d f.
\ee
In coordinates, $ (X_f)^I = -\Pi^{IJ}\pp_J f $.
The function $f$ is said the (\emph{Hamiltonian}) \emph{generator} of $X_f$. Conversely, a vector field $X\in{\frak X}^1(\Phi)$ which admits a Hamiltonian generator is said \emph{Hamiltonian}. Note, not all vector fields over $\Phi$ are Hamiltonian.

Suppose now that a Lie group $G$ acts freely\footnote{I.e. the action has no fixed points: $z^g =z$ iff $g=\mathrm{id}$.} on $\Phi$ from the right:
\be
{\sf r}: G \times \Phi \to \Phi, \quad (g, z) \mapsto z^g =  {\sf r}_g(z).
\ee
Infinitesimally, the right action ${\sf r}_g$ defines a map from elements of the Lie algebra $\frak g$ into the space of vector fields $\mathfrak{X}^1(\Phi)$:
\be
\cdot^\# : {\frak g}  \to \mathfrak{X}^1(\Phi), \quad \xi \mapsto \xi^\# := \frac{\d}{\d t}_{|t=0}  {\sf r}_{\exp (t \xi)}.
\ee

If the symplectic structure $\Omega$ is left invariant by the flow of $\xi^\#$,
\be
{\sf L}_{\xi^\#} \Omega = 0,
\ee
we say that $\xi$ is an (\emph{infinitesimal kinematical}) \emph{symmetry} 
of $(\Phi,\Omega)$. 

According to Cartan's formula, the Lie derivative ${\sf L}_X$  when applied to differential forms $\alpha\in\Omega^\bullet(\Phi)$ can be computed  as ${\sf L}_X\alpha = {\sf i}_X \d\alpha + \d {\sf i}_X\alpha$. Hence, thanks to the closedness of $\Omega$ and to Poincaré's lemma, a symmetry always admits a Hamiltonian generator $H_\xi$ at least \emph{locally} over $\Phi$
\be
{\sf i}_{\xi^\#} \Omega = - \d H_\xi.
\label{eq:HamFlow}
\ee
If a global $H_\xi:\Phi\to \bb R$ exists such that this equation holds globally over $\Phi$, then $\xi$ is said an (\emph{infinitesimal kinematical}) \emph{Hamiltonian symmetry} of  $(\Phi,\Omega)$. 

\begin{Rmk}[Invariant symplectic potential]\label{Rmk:InvSymplPot}
A sufficient condition for $\xi$ to be a Hamiltonian symmetry is that $\Omega$ admits an invariant symplectic potential, i.e. that a $\theta\in\Omega^1(\Phi)$ exists such that $\Omega = \d \theta$ and ${\sf L}_{\xi^\#} \theta = 0$. In this case, it follows from Cartan's formula that $H_\xi = {\sf i}_{\xi^\#}\theta$.
\end{Rmk}

 If all $\xi\in\frak g$ are Hamiltonian symmetries of $(\Phi,\Omega)$, then $G$ is said to be a Hamiltonian symmetry group of $(\Phi,\Omega)$.

\paragraph{Symplectic foliations}\label{sec:SymplFol}
A symplectic manifold must be even dimensional. This follows from the skew symmetry and non-degeneracy properties of $\Omega_{IJ}$. Physically, this statement codifies the idea that ``to each coordinate $q^i$ there corresponda a momentum  $p_i$." This is precisely the content of the Darboux-Weinstein theorem \cite{FernandesBookPoisson,Marsden_1999}, which states that any symplectic manifold can (locally) be equipped with coordinates $(q^i, p_i)$ such that the $\Omega$ and $\Pi$ take (locally) the form \eqref{eq:DarbouxOm} and \eqref{eq:DarbouxPi}.

So, what if we insist that $\Phi$ is odd-dimensional? Then $\Phi$ certainly cannot be a symplectic space. It can, however, still be Poisson. The question that interests us right now is: can such a $(\Phi,\Pi)$ be organized into symplectic subspaces nonetheless? Up to technicalities, the answer is yes. 

Without trying to review the general theory \cite{FernandesBookPoisson,Marsden_1999}, let's consider the following simple but important example.

\begin{figure}
\begin{center}
\includegraphics[width=5cm]{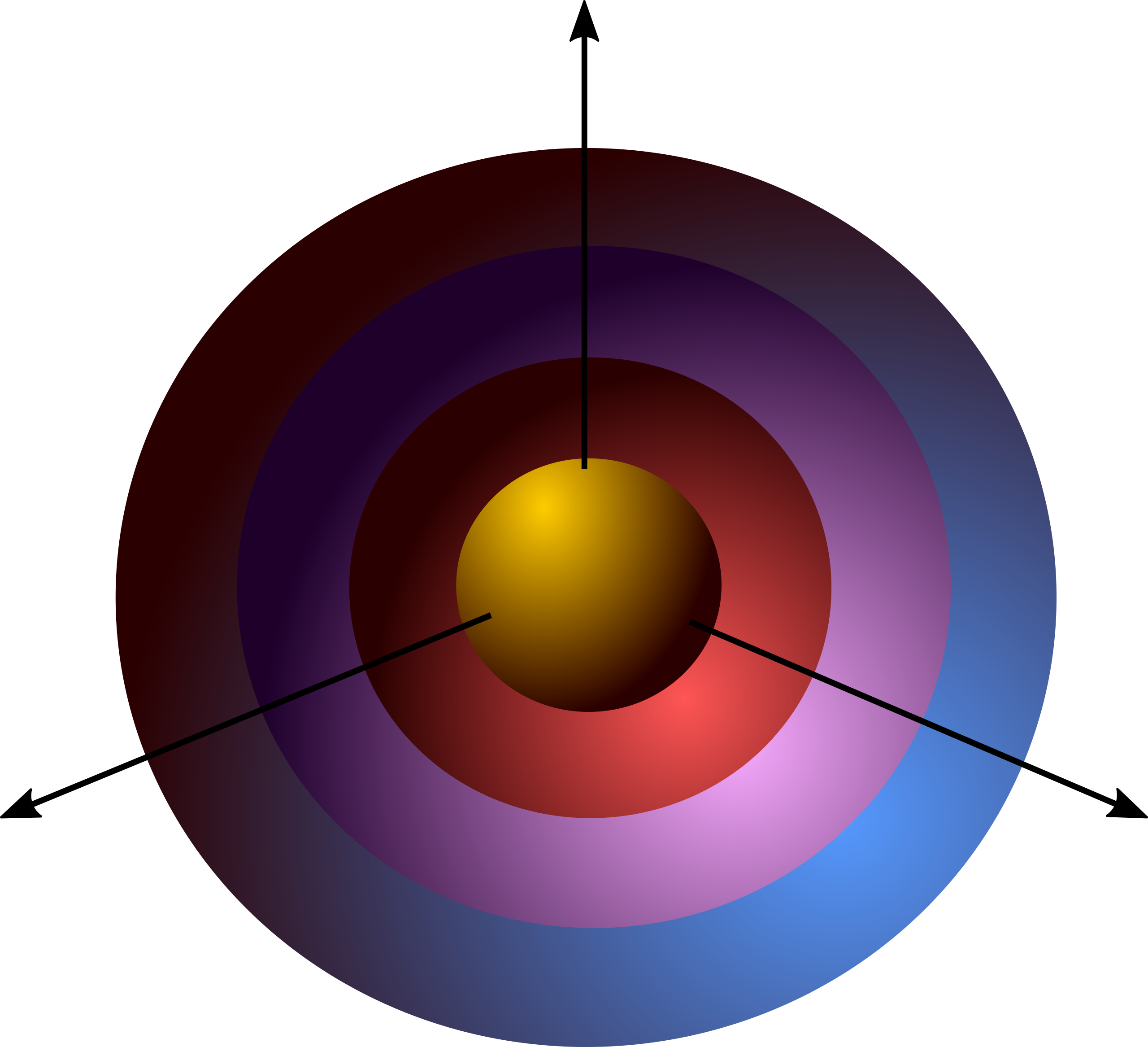}
\caption{A graphical representation of example \ref{Ex:R3}.}
\label{fig:R3}
\end{center}
\end{figure}

\begin{Ex}[Symplectic foliation of $\Phi =\bb R^3$, Figure \ref{fig:R3}]\label{Ex:R3}
Consider $\Phi= \bb R^3$, equipped with the Cartesian coordinates $\{z^I\}_{I=1,2,3}$ and the bi-vector $\Pi$,\footnote{Technically, we should rather consider $\Phi = \bb R^3\setminus\{0\}$.}
\be
\Pi^{IJ}(z) = \epsilon^{IJ}{}_Kz^K.
\ee
Then, $(\Phi,\Pi)$ is a Poisson manifold.
Being $\Phi$ odd-dimensional, the bivector $\Pi$ is necessarily degenerate, and so is its corresponding Poisson bracket.
Indeed, it is easy to check that the radius-function is in its kernel:
\be
 \{ \rho, \cdot \} \equiv 0, \qquad \rho(z) := \sqrt{\delta_{IJ} z^Iz^J}.
 \label{eq:radius}
\ee
Hence $\Pi$ is tangent to the spheres of constant radius $\rho=R$, $\bb S^2_{R}$.
Since the radius-function is the only kernel of $\{\cdot\,,\cdot\}$, $\Pi$ induces a 1-parameter family of non-degenerate bivectors $\Pi_R$ on the concentric spheres $\bb S^2_R$. These non-degenerate bivectors correspond to the 2-forms $\Omega_\rho$:
\be
\Omega_R(\theta,\phi)= R \sin\theta \wedge \d \phi \in \Omega^2(\bb S^2_R),
\ee
where we shifted to spherical coordinates, $(z^1,z^2,z^3)= (\rho\sin\theta \cos\phi, \rho\sin\theta \sin\phi, \rho\cos \theta)$.
The 2-forms $\Omega_R$ are proportional to the volume forms on the round 2-spheres, and thus they are manifestly symplectic.

Also, from the rotational invariance of $\Omega_R$ it follows that rotations around any axis $\hat \xi$ all leave $\Omega_\rho$ invariant and are therefore symmetries. Their flow is $\xi^\#= \epsilon^{IJ}{}_K z^K  \xi_I \frac{\pp}{\pp z^J}\in\mathfrak{X}^1(\Phi)$ and, with respect to $(\bb S^2_R,\Omega_R)$ they are Hamiltonian symmetries of generator
\be
H_\xi(z) = \xi_I z^I {}_{|\rho = R}.
\ee

Summarizing, on the 2-spheres $\bb S^2_{\rho=R}$, $z^3 =R\cos\theta$ and $\phi$ are conjugate Darboux coordinates, so to say ``the $q$ and the $p$.'' From the viewpoint of $\Phi=\bb R^3$, the radial coordinate $\rho$ is left on its own: it has no conjugate variable \eqref{eq:radius}. 
Therefore, $(\bb R^3,\Pi)$ is a Poisson manifold which fails to be symplectic but which is nonetheless foliated by symplectic submanifolds, the 2-spheres of constant radius $\bb S^2_{\rho=R}$. We thus call $(\Phi,\Pi)$ a \emph{symplectic foliation}, and the 2-spheres $(\bb S^2_R,\Omega_R)$ its \emph{symplectic leaves}.
\end{Ex}

Although what follows will not be needed until the end of the article---that is until the discussion of non-Abelian Yang--Mills theories in bounded regions, Section \ref{sec:YM}---we take advantage of the previous example to introduce a few more  concepts.

Indeed, far from being some artificial example of symplectic foliation, the previous example is the simplest case of a general and fully canonical construction that turns (the dual of) a Lie algebra into a Poisson space admitting its (co)adjoint orbits as symplectic leaves. 
This construction is named after Kirillov, Konstant and Sourieu (KKS), see e.g. \cite[Ch.14]{Marsden_1999}, and we will know sketch the very basics of it.

Let $\{\tau^I\}$ be a basis of $\frak g = \Lie(G)$; in this basis the structure constants of $\frak g$ are $f^{IJ}{}_K$,
\be
[ \tau^I, \tau^J]_{\frak g} = f^{IJ}{}_K \tau^K.
\ee
Let $e_I$ be the dual basis of $\frak g^*$, $\langle e_I,\tau^J\rangle = \delta^J_I$,
so that a point $z = z^I e_I \in \frak g^*$ is coordinatized by the coordinates $z^I$.
Then, one can define the following bi-vector over $\frak g^*$:
\be
\Pi = f^{IJ}{}_Kz^K  \frac{\pp}{\pp z^I}\otimes\frac{\pp}{\pp z^J}.
\ee
(This bivector can also be defined intrinsically, with no reference to a coordinate system.)
To each $\xi = \xi_I\tau^I\in \frak g$ is associated a vector field $\xi^\# = f^{IJ}{}_K z^K \xi_I \frac{\pp}{\pp z^J}$ which corresponds to the infinitesimal coadjoint action of $G$ on $\frak g^*$.
This vector field is Poisson-generated by $H_\xi = \xi_I z^I = \langle z, \xi\rangle $, $\xi^\# = \{ H_\xi, \cdot\}$, and can therefore be considered an infinitesimal symmetry of $(\frak g^*, \Pi)$.

Manifestly, each of these symmetries is tangent to the coadjoint orbits, and so is $\Pi$ itself.
Indeed, it turns out, $\Pi$ \textit{always} induces on the coadjoint orbits of $\frak g^*$ a non-degenerate bivector, i.e. a symplectic structure. 
Denoting $(S_R, \Omega_R)_R$ this symplectic foliation, Kirillov's coadjoint orbit method states\footnote{The formulas presented here are somewhat heuristic, but can be made more precise \cite{Kirillov}.} that the quantization of each $(S_R, \Omega_R)$ leads to a finite-dimensioinal Hilbert space corresponding to the irreducible $R$-th representation of $G$ of dimension $d_R \sim \tfrac{1}{2\pi} \int_{S_R} \Omega_R^{\wedge n}$, where $2n = \dim S_R$.

In the Example \ref{sec:SymplFol} right above, we implicitly considered $G=\SU(2)$, $\frak g^* \cong \bb R^3$, $f^{IJ}{}_K = \epsilon^{IJ}{}_K$, and the coadjoint action of $\SU(2)$ on $\bb R^3$ corresponds to $\SO(3)$ rotations---which explains why the coadjoint orbits of $\frak g^* \cong \bb R^3$ are 2-spheres.
Quantization of the symplectic structures $(\bb S_R^2,\Omega_R)$ then leads to the $\SU(2)$ irreducible representation of dimension $d_R \sim 2R$, from which one finds that $R$ must be a half-integer---the ``spin'' $R\equiv j$. The states in each irreducible representation are then labelled by magnetic numbers $m\in\{-R,-R+1,\dots,+R\}$ corresponding to the quantized values of the Darboux coordinate $z^3=R\cos\theta \in [-R,R]$.

\begin{Rmk}[Summary]
In this section we reviewed two ideas.
The first is that of \emph{symplectic foliation}---a space which is not symplectic itself (e.g. because it is of odd dimension) but which can be foliated by symplectic subspaces.
The second is that of the \emph{KKS construction}, which says that (the dual of) a Lie algebra admits a canonical symplectic foliation by its coadjoint orbits.
Example \ref{Ex:R3} encapsulate both ideas.
\end{Rmk}

\section{Maxwell theory without boundaries}\label{sec:MaxNoBdry}

In this section we will deal with the infinite dimensional phase space $\Phi$ of Maxwell theory over a simply connected\footnote{This affords us the freedom of not discussing Ahronov-Bohm phases.} Cauchy surface $\Sigma$ ($\pp \Sigma =\emptyset$).
We will do so formally, introducing a ``functional'' coordinate system over $\Phi$.
Since we will consider at the same time the geometry of the phase space $\Phi$ and that of the underlying space $\Sigma$, we need to ``double'' our geometric notation: one copy for the infinite dimensional $\Phi$, and one for the finite dimensional $\Sigma$.

\paragraph{Some notation}
As a rule of thumb, ``double struck'' symbols represents operations in field space, so that if $(\d, {\sf L}, {\sf i})$ are the basic operators of the finite-dimensional Cartan's calculus, $(\dd, \bb L, \bb i)$ represent analogous operators in field space. Similarly, we shall denote field-space vector fields by $\bb X\in\mathfrak{X}^1(\Phi)$.

For example, denoting $\phi^I(x)$ a set of functional coordinates over $\Phi$,
a vector field over $\Phi$, can be written as
\be
\bb X = \int_\Sigma \d^d x\, X^I(x)\frac{\delta}{\delta \phi^I(x)}  \equiv \int X^I\frac{\delta}{\delta \phi^I}
\ee
with $\delta/\delta \phi^I(x)$ a Gateaux derivative. Here, $I$ and $x$ should be thought as two different ``indices'' of our functional coordinate system, whose ranges are finite $I\in\{1,...,m\}$ and infinite $x\in\Sigma$, respectively. 

To exemplify our formal notation, let us consider a differentiable function(al) $F:\Phi\to \bb R$ and a 1-form $\alpha = \int \alpha_I \dd \phi^I \in\Omega^1(\Phi)$; then,
\be
\bb XF(\varphi) = \frac{\d}{\d t}_{|t=0} F(\phi^I + t X^I)
\ee
and
\be
\bb i_{\bb X} \alpha = \int \alpha_I X^I.
\ee
And so on.
We will not attempt to make these manipulations rigorous, since in all the cases we will need them, their meaning will be clear.

Other symbols will be introduced along the way.

\paragraph{Maxwell phase space}
In temporal gauge ($A_0 = 0$), the \emph{off-shell configuration space} of Maxwell theory\footnote{In this article, ``off-shell'' and ``on-shell'' will always refer to the Gauss constraint over $\Phi$, not to the equations of motion (which we will not consider).} is defined as the set of imaginary-valued 1-forms $A$ over the Cauchy surface $\Sigma$, i.e.\footnote{We are here assuming that the principal fibre bundle  $P\to\Sigma$ throguh which $A$ is defined, is trivial. Non-trivial bundles can be included by working in charts with \emph{fixed} transition functions. The full phase space would then be split into connected components associated with different bundle topologies. What follows can be readily adapted to each fixed such component.}
\be
\A = \Omega^1(\Sigma, \bb R) \ni A.
\ee
The \emph{off-shell Maxwell phase space} is then defined as the cotangent bundle
\be
\Phi =\T^*\A
\ee
equipped with the corresponding canonical symplectic 2-form.

The canonical coordinates over $\Phi$, analogous to the $z^I = (q^i, p_i)$ considered in the previous section, are now given by
\be
\phi^I(x) = \big( A_i(x) , E^i(x) \big).
\ee
With this notation, the canonical 1-form and 2-forms over $\Phi$ read respectively:
\be
\theta = \int_\Sigma \d^d x \sum_i E^i(x)\dd A_i(x) \equiv \int E^i \dd A_i
\ee
and, introducing $\curlywedge$ as the infinite dimensional analogue of $\wedge$,
\be
\Omega = \dd \theta = \int \dd E^i \curlywedge \dd A_i.
\ee
Comparison with the Lagrangian of the theory, shows that $E^i(x)$ corresponds to the electric field.

\paragraph{Gauge transformations}
Let $G=\mathrm{U}(1)$ be the  \emph{charge group} of Maxwell theory.
Then the associated \emph{gauge group} $\G$ is the infinite dimensional group of $G$-valued functions equipped with point-wise multiplication,
\be
\G := \Omega^0(\Sigma, \mathrm{U}(1)) \ni g,
\ee
whereas the associated Lie-algebra is\footnote{For the Abelian Maxwell theory, we adopt the hermitian convention for ${\frak g}=\Lie(G)$ so that $g\sim \exp(i \xi)$.}
\be
\Lie(\G) = \Omega^0(\Sigma, \bb R) \ni \xi.
\ee
We will call the $\xi$'s (\emph{infinitesimal}) \emph{gauge transformations}.

The action of a gauge transformation on $A$ and $E$ is $(A,E)\mapsto ( A -\mathrm i g^{-1}\d g , E)$ or, infinitesimally, $(\delta_\xi A, \delta_\xi E) = (\d \xi , 0)$.
This action defines the following vector field on $\Phi$:
\be
\xi^\# = \int \delta_\xi \phi^I \frac{\delta}{\delta \phi^I} = \int \pp_i \xi \frac{\delta}{\delta A_i}.
\ee

Since the fields $\phi^I(x)$ are nothing else than coordinate functions over $\Phi$, we write: $\bb L_{\xi^\#}\phi^I(x)  = \xi^\#\phi^I(x) = \delta_\xi \phi^I(x)$.

Note that since the $\xi$'s are mere parameters that do not depend on the physical fields, $\dd \xi \equiv 0$.
Using this fact as well as the previous formulas for the Lie derivative, one readily computes\footnote{Using Cartan's formula, it is immediate to see that $[\bb L , \dd] = 0$. Also, in the present formalism, $[\d , \dd ] =0$.}
\be
\bb L_{\xi^\#} \theta = \int E^i \pp_i \dd \xi \equiv 0.
\label{eq:gaugeinv}
\ee
Hence, from Remark \ref{Rmk:InvSymplPot}, it readily follows that $\xi^\#$ is Hamiltonian and its generator is:
\be
H_\xi = \bb i_{\xi^\#}\theta =  \int E^i \pp_i \xi
\qquad\text{and}\qquad
\bb i_{\xi^\#} \Omega = -\dd H_\xi.
\label{eq:Hxi}
\ee
\begin{Rmk}[Hamiltonian gauge symmetries]\label{Rmk:HamGaugeSym}
Notice that \emph{no} integration by parts was necessary to derive equations \eqref{eq:gaugeinv} and \eqref{eq:Hxi}.
\end{Rmk}

\begin{figure}
\begin{center}
\includegraphics[width=.3\textwidth]{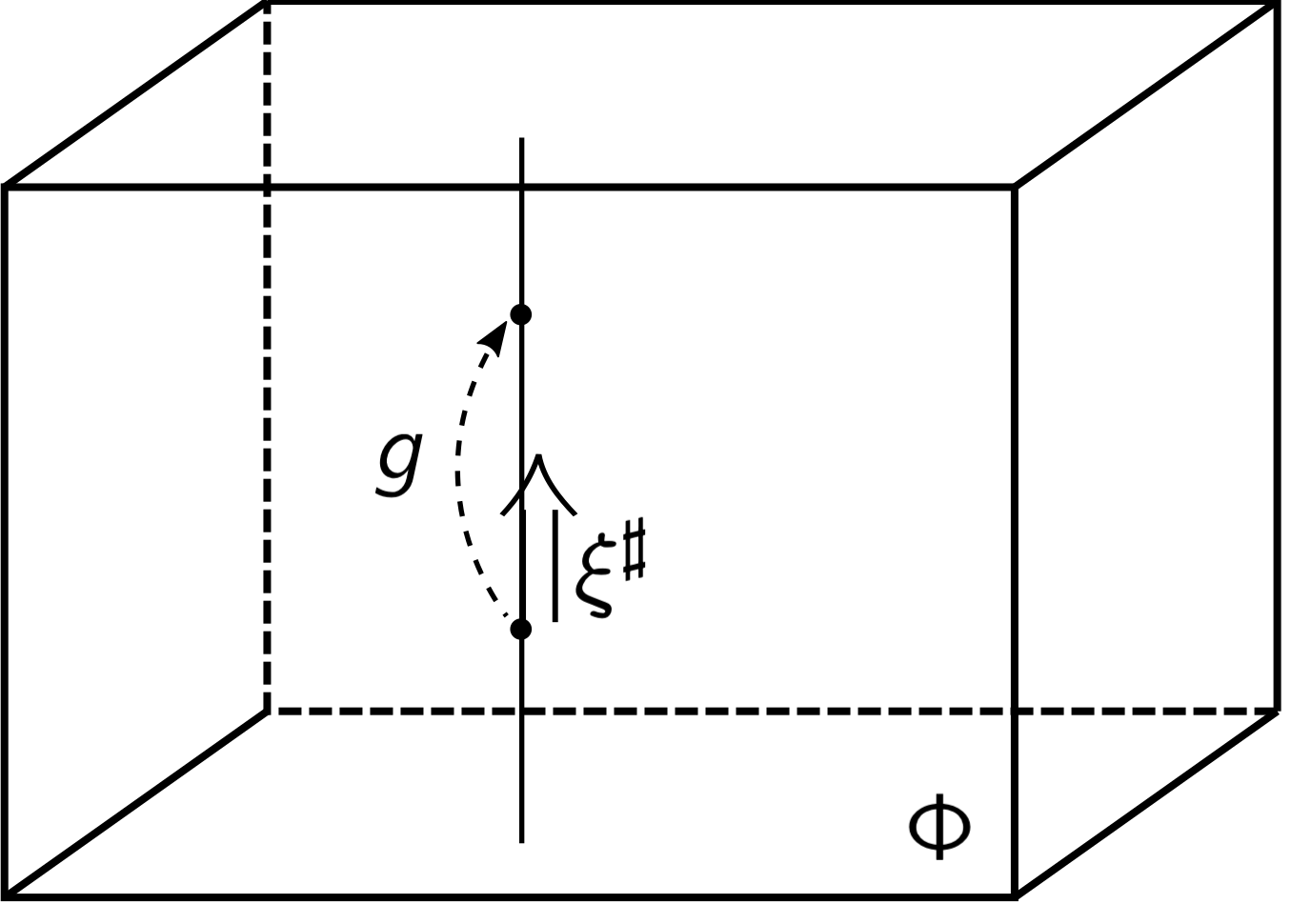}
\caption{A graphical representation of $\Phi$ and of the action of finite and infinitesimal gauge symmetries over it. The two dots represent two configurations $\varphi\in\Phi$ and $\varphi^g$ related by a finite gauge transformation $g\in\G$; the  line represents the orbit of $\G$ through the configuration $\varphi$; finally, the double arrow represents the value at $\varphi$ of the vector field $\xi^\#$ associated to an infinitesimal gauge symmetry $\xi\in\Lie(\G)$---notice that it is tangent to the orbit through $\varphi$.}
\label{fig:Phi}
\end{center}
\end{figure}

A graphical representation of $\Phi$ and gauge symmetries is provided in Figure \ref{fig:Phi}.

\paragraph{The Gauss constraint}
Not all configurations in $\T^*\A$ are however physical: only those satisfying the Gauss constraint $\GC = 0$ are,\footnote{For simplicity we will consider $R$ as a portion of a (flat) hyperplane in Minkowski space. This hypothesis is easily relaxed to general (intrinsic) geometries, with a slight complication of the notation as the only note-worthy consequence, see \cite{GomesHopfRiello,GomesRiello-quasilocal,AldoNew}.}
\be
\GC := \pp_i E^i.
\ee
Therefore we define the subspace of \emph{on-shell configurations}:
\be
\Phi_o :=\{ \phi\in\Phi : \GC(\varphi) = 0 \},
\ee
as well as the natural embedding
\be
\iota : \Phi_o \hookrightarrow \Phi.
\ee

Gauge transformations are tangent to $\Phi_o$ and, as well known, the Gauss constraint is the Hamiltomian generator of gauge transformations:
\be
- H_\xi \stackrel{\text{i.b.p.}}{=}  \int \xi\GC  =: \GC[\xi].
\ee
Here we emphasized that an integration by parts was necessary. Since $\pp\Sigma = \emptyset$, this step is warranted.

From this, we deduce that the pullback of the canonical 2-form $\Omega$ to the space of on-shell configurations $\Phi_o$ is degenerate. Indeed, evaluating \eqref{eq:Hxi} on-shell---i.e. pulling it back to $\Phi_o$---one obtains:
\be
\bb i_{\xi^\#} (\iota^*\Omega) = - \dd\iota^*  H_\xi \stackrel{\text{i.b.p.}}{=} \dd \iota^*\GC[\xi] = 0.
\label{eq:kernel}
\ee

This means that the space of on-shell configurations fails to be symplectic because $\iota^*\Omega$ has a nontrivial kernel provided by all gauge transformations.
There is however no reason to worry, because we know that $\Phi_o$ is {\it not} the ``physical'' phase space: its quotient by the action of gauge transformations is.

\begin{figure}
\begin{center}
\includegraphics[width=.30\textwidth]{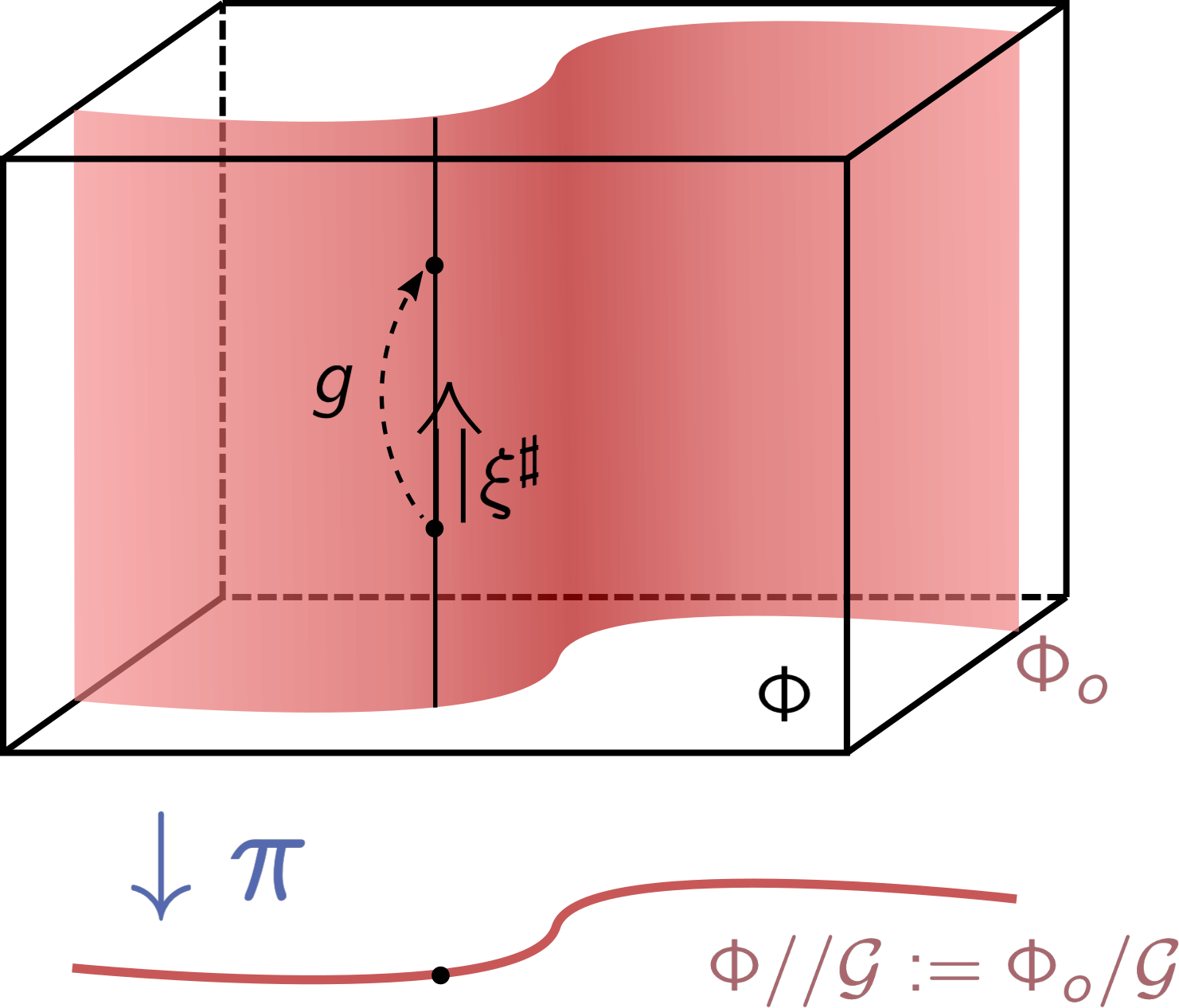}
\caption{A graphical representation of the on-shell space $\Phi_o\subset\Phi$ (the pink surface) and of the reduced phase space $\Phi//\G$ (the pink line). Gauge transformations are tangent to $\Phi_o$.}
\label{fig:PhiRed}
\end{center}
\end{figure}

See Figure \ref{fig:PhiRed}.

\paragraph{Symplectic reduction\label{sec:symplrednobdry}}
Define the \emph{reduced phase space} as the space of on-shell configurations modulo gauge transformations,
\be
\Phi//\G := \Phi_o / \G,
\ee
so that $\pi$ is the associated projection (see Figure \ref{fig:PhiRed}),
\be
\pi : \Phi_o \to \Phi//\G.
\ee

The question is then whether $\Phi//\G$ naturally carries a symplectic structure $\Omega_r$. 

Heuristically, we would like to ``project'' $\iota^*\Omega$ onto $\Phi//\G$. But forms can be naturally pulled-back, not pushed-forward. So some conditions must exist for $\iota^*\Omega$ to be ``projectable'' onto $\Phi//\G$. These conditions are precisely encapsulated by the notion of \emph{basic} forms: $\iota^*\Omega$ is basic with respect to $\pi:\Phi_o\to \Phi//\G$ if and only if it is gauge-invariant ($ \bb L_{\xi^\#}\iota^*\Omega =0$) and ``horizontal'' ($\bb i_{\xi^\#}\iota^*\Omega =0$).
More precisely, projectability can be stated as follows: if $\iota^*\Omega$ is basic, then a \emph{unique} $\Omega_r$ exists such that 
\be
\pi^*\Omega_r = \iota^*\Omega.
\ee

Moreover, if $\ker(\iota^*\Omega)= \mathrm{span}(\xi^\#)$,\footnote{Notice that the requirement of being basic implies only the inclusion $\ker(\iota^*\Omega)\supset \mathrm{span}(\xi^\#)$.} then $\ker (\Omega_r) = \{0\}$. To check that $(\Phi//\G,\Omega_r)$ is symplectic it is then enough to check that $\Omega_r $ is closed. That this is the case follows from the following facts: \i differentiation commutes with the pullback operation, \ii $\Omega$ is closed, and \iii $\pi$ is surjective.

\begin{Rmk}[Summary]\label{Rmk:conditions}
 $(\Phi,\Omega)$ induces a symplectic structure $\Omega_r$ on the reduced phase space $\Phi//\G$ if and only if the following two conditions are satisfied:
\be
\begin{cases}
\bb L_{\xi^\#}\iota^*\Omega = 0 & \text{(gauge invariance)}\\
\ker(\iota^*\Omega)= \mathrm{span}(\xi^\#) & \text{(kernel condition)}
\end{cases}
\label{eq:SymplRedSummary}
\ee%
\end{Rmk}

In the case of Maxwell theory, gauge invariance holds even off-shell and can be shown simply by acting with $\dd$ on equation \eqref{eq:gaugeinv}. Once again, no integration by parts is necessary here.

For what concerns the kernel condition, the inclusion $\ker(\iota^*\Omega)\subset \mathrm{span}(\xi^\#)$ follows directly from \eqref{eq:kernel}. The opposite inclusion also holds. Although we will not show it explicitly here, it will be a consequence of the considerations of the next section.

We conclude that, in the absence of boundaries, Maxwell theory can be seamlessly reduced onto $\Phi//\G$. I.e. $(\Phi,\Omega)$ induces a unique symplectic structure $\Omega_r$ onto the reduced phase space $\Phi//\G$.

\begin{figure}
\begin{center}
\includegraphics[width=.8\textwidth]{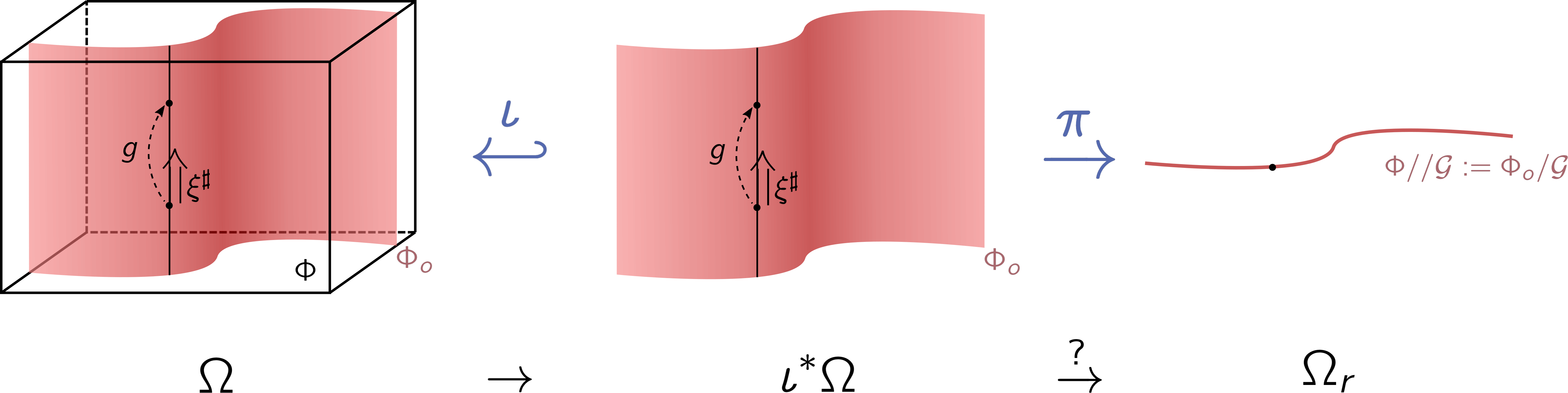}
\caption{A graphical representation of symplectic reduction. The question mark in the bottom line indicates that the ``projection'' of $\iota^*\Omega$ from the on-shell to the reduced spaces is not warranted unless conditions \eqref{eq:SymplRedSummary} are met.}
\label{fig:SymplRed}
\end{center}
\end{figure}

The procedure of symplectic reduction is represented in Figure \ref{fig:SymplRed}.

\paragraph{A closer look}\label{sec:closerlook}
The above derivation of the reduced phase space of Maxwell theory is quite abstract and does not  provide a description of the physical degrees of freedom---i.e. it does not tell us what kind of fields are contained in the reduced space $\Phi//\G$.
A closely related fact is that the above characterization of $\Omega_r$ did \textit{not} require any gauge fixing. This is a crucial point.

However, if we want to have an explicit characterization of the physical degrees of freedom, it can be convenient to resort to a gauge fixing. In this and the following sections, I will make use of the Coulomb gauge because it is highly convenient to deal with. Only in Section \ref{sec:varpi}, we will discuss how to appropriately relax not only the choice of Coulomb gauge, but also the very notion of gauge fixing. For now, let's keep things simple.

We start by splitting the gauge field into a pure-gauge and a gauge-fixed part, which I will call ``radiative,''
\be
A = A^\rad + \d \varsigma.
\ee
(Note: in this decomposition $\varsigma$ is a ``component'' of $A$, i.e. a coordinate on $\Phi$; it is \emph{not} a gauge parameter. As such $\dd \varsigma \neq 0$, whereas $\dd \xi \equiv 0$.)
In Coulomb gauge, this decomposition is by definition just the Helmholtz decomposition of the 1-tensor $A_i$, i.e.
\be
\pp^i A^\rad_i = 0 
\qquad\text{and hence}\qquad
\Delta \varsigma = \pp^i A_i.
\label{eq:Adec}
\ee

Similarly, we split the electric field into a Coulombic and radiative components:
\be
E^i = E^i_\rad + \pp^i \varphi,
\ee
such that
\be
\pp_i E_\rad^i = 0 
\qquad\text{and hence}\qquad
\Delta \varphi = \pp_i E^i = \GC
\label{eq:Edec}
\ee

A few comments are in order. 
(\textit{1}) First, the above decompositions split $A_i$ and $E^i$ into \emph{functionally independent components};  $(A^\rad, \varsigma)$ and $(E_\rad,\varphi)$ should be interpreted as a finer characterization of the coordinates $A_i$ and $E^i$ over $\Phi$.
(\textit{2}) Second, whereas $\varsigma$ is ``pure gauge'' and hence unphysical (i.e. it cannot be accessed by measurement), $\varphi$ is as physical as any other component of the electric field. Second, we see that the burden of satisfying the Gauss constraint falls all on the Coulombic part of the electric field. In particular, in the absence of matter, imposing $\GC = 0$ corresponds to fixing $\varphi = 0$:
\be
\GC = 0 \qquad\Rightarrow\qquad \varphi = 0.
\ee
And (\textit{3}) third, the Coulomb gauge decomposition for $A$ and the Coulombic/radiative decomposition for $E$ are dual to each other, in the sense that they lead to the following factorization of $\Omega$:
\be
\Omega = \int \dd E^i \curlywedge \dd A_i = \int \dd E^i_\rad  \curlywedge \dd A_i^\rad + \int \dd \pp^i \varphi \curlywedge \dd \pp_i\varsigma,
\label{eq:CRduality}
\ee
which features no mixed terms coupling $E_\rad$ with $\varsigma$, nor $A^\rad$ with $\varphi$.

Now, from equation \eqref{eq:CRduality}, integrating by parts, we can isolate the Gauss constraint:
\be
\Omega\stackrel{\text{i.b.p.}}{=} \int \dd E^i_\rad  \curlywedge \dd A_i^\rad - \int \dd \GC \curlywedge \dd \varsigma.
\label{eq:OmegaGauss}
\ee
From this expression we can deduce that the pure-gauge part of $A$ is the degree of freedom conjugate to the Gauss constraint.
Since momenta always generate translations in the conjugate coordinate, this is nothing else than a restatement of the fact that $\GC$ generates gauge transformations.

But this expression also offers a very simple interpretation of symplectic reduction: as a consequence of the imposition of the Gauss constraint $\GC=0$, i.e. of the pullback of $\Omega$ to $\Phi_o$, the pure-gauge degree of freedom $\varsigma$ ends up lacking a symplectic partner: modding out gauge transformations basically means getting rid of $\varsigma$ as well. By removing \emph{an entire symplectic pair}, symplectic reduction turns the ``larger'' symplectic 2-form $\Omega$ into a ``smaller'' symplectic 2-form $\Omega_r$, without jeopardizing in the process its symplectic nature---and in particular its non-degeneracy.

Thus, up to the pullback $\pi^*$, we can think of $\Omega_r$ as the radiative part of $\Omega$, since
\be
\pi^*\Omega_r = \int \dd E^i_\rad  \curlywedge \dd A_i^\rad .
\label{eq:OmegaRad}
\ee

We conclude our discussion of symplectic reduction of Maxwell theory in the absence of boundaries with a few remarks.

\begin{Rmk}[The radiative electric field]\label{Rmk:Erad}
Since the pure-gauge part of $A$ is always of the form $\d \varsigma$, one can \emph{canonically define} radiative modes of $E$ through the duality condition \eqref{eq:CRduality}, that is $\int \dd E^i_\rad \curlywedge \pp_i \dd \varsigma \equiv 0$. In this way, not only the radiative modes of $E$ are canonically divergence-free but they will also \emph{always} drop from the Gauss constraint---irrespectively of any gauge choice for $A_\rad$. In other words, the functional properties of $E_\rad$ are given a priori, in the same way as the pure-gauge prat of $A$ is always a pure-gradient.

In turn, the choice of gauge condition fixing the functional form of $A_\rad$ will determine what kind of functional form the Coulombic part of $E$ will take, through the second duality condition $\int \dd E^i_\Coul \curlywedge \dd A^i_\rad \equiv 0$. The decomposition of a given configuration $(A,E)$ into its radiative and pure-gauge/Coulombic components can be performed only once a choice of gauge for $A^\rad$ has been made; two different such choices lead to different definitions of the Coulombic component of $E$ which always differ by a radiative mode. In the above example, the decomposition turns out to be self-dual. This is one reason why it is particularly nice to work with.

 One useful way to think about different choices of the radiative/pure-gauge decompositions is in terms of choices of ``(functional-)coordinate axes" on $\A$: a canonical one along the pure-gauge direction and an arbitrary one transverse to it.\footnote{As we will explain in Section \ref{sec:YM}, it is actually best to think of these as choices of coordinate axes on $\T\A$, rather than $\A$ itself. This change of perspective is unavoidable in the non-Abelian theory.} As usual in a geometrical setup, although a choice of coordinates has to be made in order to have an explicit description of the space at hand, the precise choice one commits to is ultimately irrelevant. This can be explicitly shown through the formalism introduced in section \ref{sec:YM}, which allows us to leave these choices of coordinates completely arbitrary. In the meantime, we shall keep working with the Coulombic decomposition, while being reassured that $\Omega_r$ is well-defined independently of this choice thanks to abstract and coordinate-free arguments such as those of Section \ref{sec:symplrednobdry}.

In this regard, notice that since any two choices of gauge necessarily lead to definitions of the Coulombic electric field that differ by a radiative electric field (whose functional properties are defined a priori), the space of radiative electric fields \emph{modulo gauge} is like an affine space whose ``origin'' can be arbitrarily fixed---in this case by a choice of gauge. Since $\pi^*\Omega_r$ is insensitive to such a choice of ``origin'' (it depends only on difference $\dd E_\rad$) the whole treatment is indeed consistent.

These conclusions will not be affected by the introduction of boundaries.
\end{Rmk}

\begin{Rmk}[Symplectic reduction]\label{Rmk:SymplRed}
Summarizing, $\varsigma$, i.e. the pure-gauge part of $A$, is conjugate to the Coulombic part of $E$ entering the Gauss constraint, i.e. $\varphi$. Symplectic reduction is therefore a systematic procedure to get rid of this symplectic \emph{pair}: the imposition of the Gauss constraint fixes $\varphi$, forcing it to vanish, whereas $\varsigma$ is eliminated by ``modding out gauge transformations.'' The resulting symplectic structure, in essence the reduced symplectic structure, is thus composed by the radiative degrees of freedom only.
\end{Rmk}

\begin{Rmk}[Non-locality]\label{Rmk:nonlocality}
Given a field configuration $(A,E)$, the radiative/Coulombic decomposition introduced in this section is \emph{non-local}. This is because it relies on the solution of certain elliptic differential equations. In the above example, which is self-dual, these elliptic equations are both Laplace equations, \eqref{eq:Adec} and \eqref{eq:Edec}.
This is most clearly seen by re-expressing $\pi^*\Omega_r$ in terms of local fields:
\be
\pi^*\Omega_r = \iint \d^dx \d^d y \; (P_{\perp i}{}^j)(x,y) \Big( \dd E^i (x) \curlywedge \dd A_j(y)\Big),
\label{eq:OmegaPperp}
\ee
where
\be
P_{\perp i}{}^j  = \delta_i{}^j - \frac{\pp_i \pp^j}{\Delta}
\ee
is the non-local integral kernel that projects a 1-tensor over $\Sigma$ onto its transverse part.
\end{Rmk}

\begin{Rmk}[Dirac's formalism and gauge-fixings]
This discussion of the symplectic reduction is perfectly compatible with the Dirac bracket formalism.
To see this, let $\mathsf f = \pp^i A_i$ be the second-class gauge-fixing constraint that fixes the Coulomb gauge. Introducing the constraint ``vector'' $\chi^a(x) = ( \GC(x), \mathsf f(x))$, one introduces the Dirac ``matrix'' of constraints as $M^{ab}(x,y) = \{ \chi^a(x) , \chi^b(y)\}$ and thus defines the Dirac bracket as 
\be
\{\cdot\,,\cdot\}_D =  \{\cdot\,,\cdot\} - \{\cdot\,,\chi^a\}(M^{-1})_{ab} \{\chi^b,\cdot\}.
\ee
In the present case, $M^{ab} = \epsilon^{ab} \times\{ \mathsf f, \GC \} =  \epsilon^{ab} \Delta$, and one obtains\footnote{That is, $\{E_i(x), A^j(y)\}_D = (P_{\perp i}{}^j)(x,y)$. Cf. \eqref{eq:EBbracket} below.}
\be
\{\cdot\,, \cdot\}_D =\iint (P_{\perp i}{}^j)(x,y)\frac{\delta}{\delta E_i(x)} \otimes \frac{\delta}{\delta A^j(y)}.
\label{eq:DiracBr}
\ee

But we could follow the symplectic reduction formalism from even closer: indeed, fixing the coordinate function $\varsigma:= \Delta^{-1}\pp^iA_i=0$ corresponds to fixing Coulomb gauge too. However, with respect to this new gauge fixing constraint the Dirac ``matrix'' reads simply $M^{ab} = \epsilon^{ab}$---which corresponds to $\varsigma$ and $\GC$ being conjugate as in \eqref{eq:OmegaGauss}. Of course, this leads to the same Dirac bracket as before.

Finally, the fact that the Dirac bracket $\{\cdot\,,\cdot\}_D$ \eqref{eq:DiracBr} and the reduced symplectic structure $\Omega_r$ \eqref{eq:OmegaPperp} do not look like the ``inverse'' of each other \eqref{eq:PoissonBracket} despite $\Phi//\G$ being symplectic, is just an optical illusion: indeed, a projector like $P_\perp$ is the inverse of itself once restricted to its image. In other words the problem is that we have been forced to write $\{\cdot\,,\cdot\}_D$ and $\Omega_r$ over too large a space, that is over $\Phi_o$ rather than over $\Phi//\G=\Phi_o/\G$. Taking this into account, one realizes that the push-forward of the Dirac bracket's bivector \eqref{eq:DiracBr} by $\pi_*$ is indeed the ``inverse'' of the 2-form $\Omega_r$.
\end{Rmk}

\begin{Rmk}[Non-locality, again]
In 4d electromagnetism over a simply connected space, an obvious complete, gauge-invariant, and local set of field variables exists: the electric and magnetic fields $(E,B)$. Then, in which sense is Maxwell theory nonlocal? The nonlocality can be detected by looking at the symplectic in these variables. Indeed, in the absence of charges,\footnote{Notice that this bracket requires $E$ to be divergence-free. i.e. it is valid on-shell of Gauss and in the absence of charges.} the Poisson bracket between $E$ and $B$ is
\be
\{ E^i(x), B^j(y) \}_r = \epsilon^{ijk}\pp^{(y)}_k \delta(x-y)
\label{eq:EBbracket}
\ee
and therefore $\Omega_r$---which is morally speaking the inverse of $\{\cdot,\cdot\}_r$---is a nonlocal quantity. This means that in the following formula, $k_{ij}(x,y)$ is an integral kernel:
\be
\Omega_r = \iint E^i(x) k_{ij}(x,y) B^j(y).
\ee
Of course a viable choice of $k_{ij}$ is the convolution of the kernel that allows one to invert for $A_\rad$ from the equation $B = \mathrm{curl} A_\rad$ together with the $P_\perp$ above. 

A much simpler way to see that Maxwell theory is nonlocal is to resort to charged matter. Indeed, even without looking at the symplectic structure, the Gauss constraint implies that a charge particle must come equipped with an extended electrostatic field that permeates $\Sigma$. In field-theoretic terms, this means that the charged matter field is ``dressed'' (see Section \ref{sec:MaxMatt}). In this dressed-field description there are no gauge-variant quantities involved, and the origin of the nonlocality can be fully attributed to the Gauss constraint.
\end{Rmk}

\section{Maxwell theory with boundaries}\label{sec:MaxBdry}
This rather long review leaves us in a good position to easily identify what goes wrong in the presence of boundaries.

Henceforth, we will replace the Cauchy surface $\Sigma$ by $R$, a subregion of $\Sigma$ with $\pp R \neq \emptyset$.
What we will have in mind is a disk $R\cong \mathbb B^d$ with boundary $\pp R \cong \bb S^{d-1}$.

In analogy with the previous section, we define the off-shell phase space $(\Phi,\Omega)$ of Maxwell theory with boundary as
\be
\Phi = \T^*\A, \qquad \A = \Omega^1(R, \mathrm i\bb R)
\ee
equipped with the canonical 1- and 2-forms
\be
\theta = \int E^i \dd A_i \qquad\text{and}\qquad\Omega = \int \dd E^i \curlywedge \dd A_i.
\ee
Hereafter, $\int \equiv \int_R$ rather than over $\Sigma$.

For the future convenience, we introduce here the outgoing co-normal $s_i$ at $\pp R$ as well as the \emph{electric flux $f$ through $\pp R$}, i.e.
\be
f:= s_i E^i{}_{|\pp R}.
\label{eq:flux}
\ee

Finally, we define gauge transformations to be given by
\be
\G = \Omega^0(R, \mathrm{U}(1)) \ni g
\qquad\text{and}\qquad
\Lie(\G) = \Omega^0(R, \mathrm i \bb R) \ni \xi.
\ee
Notice that we demand \emph{no} restriction in the behaviour of the gauge transformations at $\pp R$: gauge will be treated as such in the bulk \emph{as well as} at the boundary. Insisting on this point is at the core of our treatment (cf. point \textit{\ref{iii}} in the Introduction).

\paragraph{Obstructions to projectability} 
Let's start by asking what, if anything, obstructs the projectability of $\Omega$ onto $\Phi//\G$ when  boundaries are present.
Recall that $\Omega$ is projectable if and only if, on shell of the Gauss constraint, it is  gauge invariance \textit{and} satisfies the kernel property \eqref{eq:SymplRedSummary}.

First we notice that, thanks to Remark \ref{Rmk:HamGaugeSym}, the gauge invariance of $\theta$ remains valid even in the presence of boundaries. In other words, even if $\pp R \neq \emptyset$, not only
\be
\bb i_{\xi^\#} \Omega = - \dd H_\xi 
\qquad\text{for}\qquad
H_\xi =  \bb i_{\xi^\#} \theta = \int E^i\pp_i \xi,
\ee
but also
\be
\bb L_{\xi^\#}\Omega = 0.
\label{eq:gauge-inv-ppR}
\ee
Therefore, even in the presence of boundaries, gauge transformations are Hamiltonian and $\Omega$ \emph{is gauge invariant}.

Hence, if anything goes wrong in the symplectic reduction procedure, this is the kernel property. Indeed, since 
\be
H_\xi \stackrel{\text{i.b.p.}}{=} - \GC[\xi]+\oint \xi f,
\ee
we see that the generator $H_\xi $ fails to vanish on-shell of the Gauss constraint and thus:
\be
\bb i_{\xi^\#} \iota^*\Omega = - \oint \xi \dd f  \neq 0 
\qquad\text{if}\qquad 
\xi{}_{|\pp R} \neq 0.
\label{eq:obstruction}
\ee
This leads us to the following:

\begin{Rmk}[Obstruction to symplectic reduction]
The obstruction posed by boundaries to a straightforward symplectic reduction (cf. \eqref{eq:SymplRedSummary})
is due to the fact that gauge-transformations that fail to vanish at the boundary also fail to be in the kernel of the on-shell 2-form $\iota^*\Omega$.
\end{Rmk}

\begin{Rmk}[Gauge invariance]
Contrary to what is sometimes claimed in the literature, the gauge invariance of $\theta$ and $\Omega$ (cf. \eqref{eq:SymplRedSummary}) is on the other hand \emph{not} jeopardized by the presence of boundaries \eqref{eq:gauge-inv-ppR}.\footnote{If we were to generalize the mathematical setup presented here  including so-called ``\emph{field-dependent} gauge transformations''---i.e. by allowing one to choose different gauge ``parameters'' $\xi$ at different configurations $A\in\A$ so that $\dd \xi\not\equiv0$ (cf. \cite[Sec.2]{GomesRiello-quasilocal} for a mathematical formalization in terms of action Lie algebroid)---then $\bb L_{\xi^\#} \iota^*\Omega = \oint \dd f \curlywedge \dd \xi \not\equiv0$. That is, $\iota^*\Omega$ fails to be gauge-invariant under this enlarged group of field-dependent gauge transformations. We notice that, in this enlarged framework, the gauge-invariance condition $\bb L_{\xi^\#}\iota^*\Omega = 0$ for all \emph{field-dependent} $\xi$---together with $\dd\Omega=0$ and the fact that $\Phi$ is affine---turns out to automatically imply the relation $\ker(\iota^*\Omega) \supset \mathrm{span}(\xi^\#)$. In other words, if $\iota^*\Omega$ is both closed and gauge invariant under all \emph{field-dependent} gauge transformations, then it is necessarily basic and hence projectable. Indeed, if $0 = \bb L_{\xi^\#}\iota^*\Omega = \dd (\bb i_{\xi^\#}\iota^*\Omega)$ then, by Poincaré Lemma and $\bb R$-linearity in $\xi$, $\bb i_{\xi^\#}\iota^*\Omega = \dd \langle \alpha , \xi\rangle$ for some $\alpha\in\Omega^0(\Phi,\mathfrak g^*)$; but, since on the other hand the contraction $\bb i_{\xi^\#}\iota^*\Omega$ manifestly does not depend on $\dd\xi$ we conclude that $\alpha = 0$ and hence $\bb i_{\xi^\#}\iota^*\Omega=0$ (this is a consequence of linearity of $\bb i_{\xi^\#}\iota^*\Omega$ under field-dependent \emph{spatially-constant} rescalings of $\xi$, i.e. under $\xi \mapsto c(A)\xi$ where $c:\Phi\to \bb R$).  Field-dependent gauge-transformations were put front and forward in previous work of ours (e.g. \cite{AldoNew}) due to the power and flexibility of the ensuing formalism. In this article, however, we opted for a more ``minimal'' approach.}
\end{Rmk}

\begin{Rmk}[Maxwell \& Yang--Mills VS. Chern--Simons \& $BF$]\label{Rmk:CSBF}
Note that in Chern--Simons theories, not only $\Phi$ fails to be a cotangent bundle equipped with canonical one- and two-forms, but \emph{any} 1-form $\theta_\text{CS}$ such that $ \dd \theta_\text{CS}=\Omega_\text{CS}: = \int \tr(\dd A \wedge \dd A)$ fails, in the presence of boundaries, to be gauge-invariant---even on-shell ($F = 0$).\footnote{And without any reference to the field-dependent gauge-transformations mentioned in the previous footnote.} The latter statement holds true also in $BF$ theories. Given the central role that $\theta_\text{CS}$ plays in quantization (e.g. \cite{bates1997}), this observation provides yet another reason not to conflate the treatment of gauge and boundaries in Yang--Mills theory with that in Chern--Simons and $BF$ theories (a more basic difference between these theories is that only Yang--Mills has a gauge-invariant Lagrangian \emph{density}).
\end{Rmk}

\paragraph{Coulombic and radiative modes}
To gain a more concrete grasp of the problem of symplectic reduction in the presence of boundary along the lines of Section \ref{sec:closerlook}, we need to acquire more familiarity with the nature of the radiative/Coulombic decomposition---and of the constraint---when boundaries are present. This analysis will naturally lead us to an in-depth discussion of the role of the electric flux $f$. 

As emphasized in Remark \ref{Rmk:nonlocality}, the radiative/Coulombic decompositions of $(A,E)$ relies on the solution of an elliptic equation, that is a Laplace equation. However, in the presence of boundaries, for the solutions of these equations to be unique, one needs to impose boundary conditions to these elliptic equations---turning them into \emph{elliptic boundary value problems} (EBVP).

To study which boundary conditions one should impose, it is instructive to start from the decomposition of $E$. 
As emphasized in Remark \ref{Rmk:Erad}, one can define the properties of $E_\rad$ with no reference to a gauge choice, but entirely from a requirement of duality (more precisely, $L^2$-orthogonality) with respect to pure-gauge transformations:
\be
\int \dd E^i_\rad \curlywedge \pp_i \dd \varsigma \equiv 0.
\ee
Now, since we insist on \emph{not} imposing any restriction on gauge transformations at the boundary $\pp R$, we see that the above duality requirement is satisfied for all $\dd \varsigma$ if and only if
\be
\begin{cases}
\pp_i E^i_\rad = 0 & \text{in }R\\
s_i E^i_\rad = 0 & \text{at }\pp R\\
\end{cases}
\label{eq:Erad}
\ee

These two equations, together with the choice of writing the Coulombic part of $E$ as a pure gradient, are enough to completely fix the decomposition. Indeed, by contracting $E^i = E^i_\rad + \pp^i \varphi$ with $\pp_i$ and $s_i$, we find the following EBVP which \emph{uniquely} fixes $\varphi=\varphi(E)$ and thus $E^i_\rad$ as well:
\be
\begin{cases}
\Delta \varphi = \pp_i E^i & \text{in }R\\
s_i\pp^i \varphi = s_i E^i & \text{at }\pp R\\
\end{cases}
\label{eq:EBVPvarphi}
\ee

Now that we have the radiative/Coulombic decomposition of $E_\rad^i$, by the duality relation $\int \dd \pp_i\varphi\curlywedge\dd A_i^\rad \equiv 0$  we can also deduce the properties of $A_i^\rad$ and hence fix the radiative/pure-gauge decomposition of $A$ in the Coulomb gauge with boundaries. Unsurprisingly, the result is completely self-dual:
\be
\begin{cases}
\pp^i A_i^\rad = 0 & \text{in }R\\
s^i A_i^\rad = 0 & \text{at }\pp R\\
\end{cases}
\qquad\text{and}\qquad
\begin{cases}
\Delta \varsigma = \pp^i A_i & \text{in }R\\
s^i\pp_i \varsigma = s^i A_i & \text{at }\pp R\\
\end{cases}
\label{eq:dressing}
\ee

Notice that a field-\emph{in}dependent (Dirichlet) boundary condition is imposed on $A_i^\rad$, whereas $\varsigma$ satisfies a field-\emph{dependent} (Neumann) boundary condition.
This specific assignment of field-dependent versus field-independent boundary conditions is necessary in order not to restrict gauge freedom at $\pp R$.\footnote{Instead, the choice of a Dirichlet vs. Neumann boundary condition for the radiative and pure-gauge parts of $A$ reflects our choice of using the Hodge-Helmholtz decomposition to characterize the radiative/pure-gauge decomposition. This is the most convenient choice we can think of, but as discussed in the following Remarks (\ref{Rmk:Uniqueness}, \ref{Rmk:fcoord}, and \ref{Rmk:CoulErmk48}) and more generally in Section \ref{sec:YM}, there is nothing truly fundamental to it.}
(Conversely, this means that any $A\in\A$ is gauge-related to a unique divergence- and flux-less configuration $A_\rad$---provided gauge freedom is left \emph{un}restricted at $\pp R$.)

\begin{Rmk}[Uniqueness of the decomposition and the electric charge]\label{Rmk:Uniqueness}
To be more precise, what is uniquely determined by these EBVP is $\pp^i\varphi$ and $\pp_i\varsigma$, but not $\varphi$ and $\varsigma$ themselves.
Indeed, because of the Neumann boundary conditions, $\varphi$ and $\varsigma$ are  only determined up to a constant. 
Importantly, the radiative/Coulombic decomposition only relies on $\pp^i\varphi$ and $\pp_i\varsigma$, and therefore this ambiguity is not an issue.
However, notice that in the case of $\varsigma$, the residual freedom we have just identified corresponds to ``global gauge transformations''---which are associated to the total electric charge contained in $R$. Here, we limit ourselves to state that this is no coincidence, and refer the reader to Section \ref{sec:MaxMatt} for a quick sketch of the consequences of this ambiguity in the presence of matter, and to \cite{GomesRiello-quasilocal} for more details.\footnote{The gist of the argument is the following. Global gauge transformations are in the kernel of this EBVP because they leave $A$ unchanged, and as such they play a similar role Killing symmetries play in general relativity. The analogy works best in the non-Abelian case, where only special configurations admit such symmetries, just like it happens in general relativity. However, the fact that all configurations in electromagnetism possess such a ``Killing symmetry'' provides the geometrical reason why a total charge can be unambiguously defined in electrogmagnetism, but not in QCD.\label{fnt:killing}}
\end{Rmk}

Consider now the EBVP \eqref{eq:EBVPvarphi} that fixes $\varphi=\varphi(E)$.
As before, its bulk part corresponds to the Gauss constraint. 
But now, this is complemented by a boundary condition---necessary to uniquely determine $\varphi$ throughout $R$---which involves precisely the flux $f$.

This fact together with equation \eqref{eq:Erad} leads us to the two following crucial observations: in the presence of boundaries and \emph{on-shell of the Gauss constraint} , \i $\varphi$ does not vanish and is instead \emph{entirely determined throughout $R$ by the value of the flux $f$}, whereas---conversely---\ii \emph{$E_\rad$ is functionally independent of $f$.}

\begin{Rmk}[Beyond Coulomb gauge]\label{Rmk:fcoord}
Although the functional form of the Coulombic part of $E$ and the type of EBVP fixing it do depend on a choice of gauge fixing, the fact that the Coulombic part of $E$ is on-shell fully determined by $f$ whereas $E_\rad$ is fully independent from it, are completely general features that only descend from \eqref{eq:Erad} which is, as we argued, completely universal.
From the point of view of $\Phi_o$ different choices of decompositon, correspond to different choices of coordinates over the on-shell configurations. Which precise coordinate the flux $f$ represents, however, depends on the chosen form of the decomposition.
\end{Rmk}

\paragraph{Symplectic foliation and superselection sectors}\label{sec:SSS}
After these clarifications on the role of the flux $f$, we can now go back and re-assess the equation \eqref{eq:obstruction}, that we report here:
\be
\bb i_{\xi^\#} \iota^*\Omega = - \oint \xi \dd f .
\label{eq:obstruction-bis}
\ee
This equation identifies the obstruction to the symplectic reduction in the boundary term $\oint \xi \dd f$. 

A first way to get around it is to declare, loosely speaking, that ``boundary gauge transformations should be treated differently,'' e.g. by reducing $\Phi$ only by bulk-supported gauge transformations. This modified reduction procedure would then proceed unhindered, but would lead to a reduced phase space larger that $\Phi//\G$. We will comment on this option, and its drawbacks, in Remark \ref{Rmk:edge} below. 

Here we shall instead focus on a different option: even if $\Omega$ cannot be directly reduced to $\Phi//\G$, does it inform a ``symplectic-like'' structure on that space?

The key idea here is to shift our attention from $\xi$ to $f$, that is: what if the culprit were not $\xi_{|\pp R}$ but $\dd f$?
If we were to restrict $\iota^*\Omega$ to submanifolds of fixed $f$, within each submanifold the reduction procedure would go through flawlessly!

Thus, define
\be
\Phi^f_o = \{ \varphi \in \Phi_o : s^iE_i{}_{|\pp R} = f \}
\ee
as well as the natural embedding 
\be
\iota_f : \Phi^f_o \hookrightarrow \Phi.
\ee
From this and the definitions of the previous section, it follows that $\iota^*\Omega  =  \int \dd E^i_\rad \curlywedge \dd A_i^\rad + \oint \dd f \curlywedge \dd \varsigma$, and therefore
\be
\iota_f^*\Omega =  \int \dd E^i_\rad \curlywedge \dd A_i^\rad .
\label{eq:iOmegaf}
\ee
Notice that $f$ has completely dropped from the right-hand-side of this formula.
Hence, from the fact that $\varphi$ is completely fixed by $f$ through the Gauss constraint, one can deduce that the only kernel of $\iota_f^*\Omega$ is given by gauge transformations.
Moreover, since \emph{in Maxwell theory $f$ is gauge invariant}, it also immediately follows that $\iota_f^*\Omega$ is gauge invariant itself. 

Summarizing, $\iota_f^*\Omega$ is found to satisfy the following two properties:
\be
\begin{cases}
\bb L_{\xi^\#}  \iota_f^*\Omega = 0 & \text{(gauge invariance)}\\
\ker(\iota_f^*\Omega ) = \mathrm{span}(\xi^\#) & \text{(kenrel property)}
\end{cases}
\ee
which are the necessary and sufficient conditions for $\iota_f^*\Omega$ to define a unique symplectic structure $\Omega^f_r$ on $\Phi_o^f/\G$ such that (cf. Remark \ref{Rmk:conditions})
\be
\pi^*\Omega^f_r = \iota_f^*\Omega.
\label{eq:Omegafr}
\ee
Finally, it is manifest that the set of subspaces $\{\Phi_f\}_f$  foliates $\Phi_o$, and that consequently the set of subspaces $\{\Phi_o^f/\G\}_f$ foliates $\Phi//\G$.

Denoting
\be
\Phi\slashf\G := \Phi_o^f/\G,
\ee
we hence conclude with first important result: \emph{in the presence of boundaries, $\Phi//\G$ naturally admits a symplectic foliation whose leaves are the sectors $\{\Phi\slashf\G\}_f$ of fixed Coulombic electric field $E^i_\Coul = \pp^i\varphi(f)$.}

\begin{figure}
\begin{center}
\includegraphics[width=.7\textwidth]{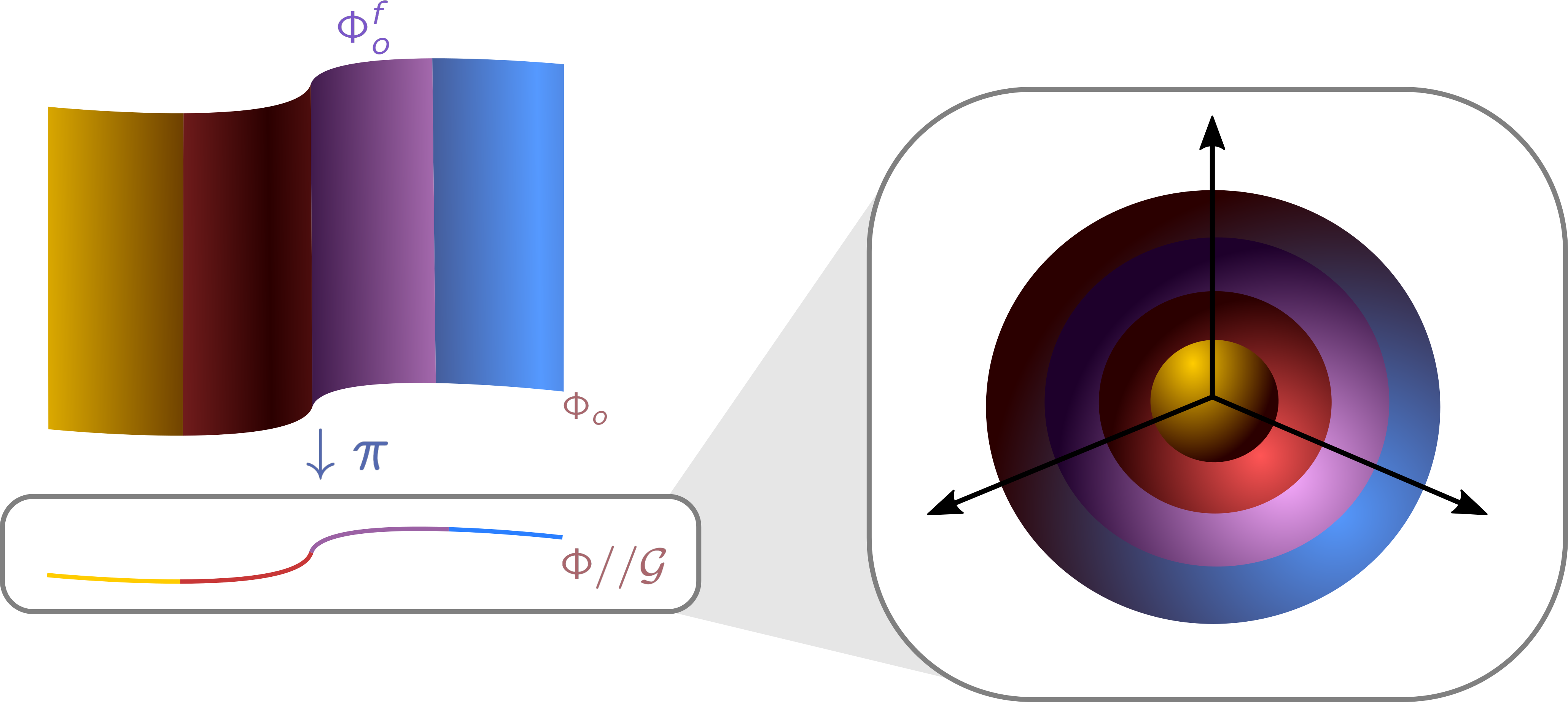}
\caption{A graphical representation of symplectic reduction in the presence of boundaries. \emph{On the left}: The on-shell space $\Phi_o$ is foliated by subspaces $\Phi_o^f$ characterized by a fixed flux $f$ (the coloured vertical strips); upon reduction, one obtains a symplectic foliation of $\Phi//\G$ by the superselection sectors $\Phi\slashf\G$ (the coloured segments). \emph{On the right}: The reduced space $\Phi//\G$ is ``blown up'' and represented as the space $\bb R^3$ equipped with the concentric-spheres symplectic foliation discussed in Example \ref{Ex:R3} and Figure \ref{fig:R3}; in this representation, every coloured sphere embodies a flux superselection sector $\Phi\slashf\G$.}
\label{fig:SSSred}
\end{center}
\end{figure}

We thus refer the reader to Figure \ref{fig:SSSred}, and end this section with a series of remarks.

\begin{Rmk}[Superselection sectors]
In physics parlance, the symplectic leaves $\{\Phi\slashf\G\}_f$ are called \emph{flux superselection sectors}, since in each of them $f$ is fixed and Poisson-commutes with all other observables. 
As emphasized in the above conclusions---in vacuum---fixing $f$ means indeed fixing the Coulombic electric field $E^i_\Coul = \pp^i\varphi(f)$ \emph{throughout} $R$, and not just at the boundary $\pp R$.
\end{Rmk}

\begin{Rmk}[Symplectic reduction]
Within each superselection sector $\Phi\slashf\G$, the symplectic reduction proceeds as in the boundary-less case (cf. Remark \ref{Rmk:SymplRed}): the  Gauss constraint completely fixes the Coulombic part of the electric field $\varphi$ (albeit not to zero) whereas the act of ``modding out'' gauge transformations gets rid of the	dof conjugate to $\varphi$. Together, these two steps get rid of an entire canonical pair of degrees of freedom, thus leaving us with a symplectic form. The remaining degrees of freedom in $\Omega_r^f$ are the radiative dof, which are functionally independent of $f$. (See Remark \ref{Rmk:Erad}.)
\end{Rmk}

\begin{Rmk}[The Coulombic electric field]\label{Rmk:CoulErmk48} 
The fact that $E_\Coul$ drops from $\pi^*\Omega^f_r$ it is not because $E_\Coul$ vanishes or is not accessible by experiments, but simply because it is fixed within each superselection sector. 
Since different choices of radiative/Coulombic decomposition lead to distinct forms of the Coulombic electric field which differ by a radiative contribution, any choice of decomposition leads, at the end of the day, to the same symplectic spaces $(\Phi\slashf\G,\Omega_r^f)$---albeit expressed in different functional coordinates. 
We refer to Remark \ref{Rmk:Erad} for more on this point, where the choice of radiative/Coulombic decomposition is compared to a choice of origin in an affine space.
\end{Rmk}

In the next section we will spend a few words on the physical interpretation of superselection sectors. 
But first we conclude this section with the promised remark on the symplectic reduction by bulk-supported gauge transformations only:

\begin{Rmk}[Edge modes]\label{Rmk:edge}
The same equations that show that $\Phi$ cannot be gauge-reduced with respect to the action of $\G$, they also show that it can be reduced with respect to the action of the group of bulk-supported gauge transformations $\mathring\G = \{ g\in\G: g{}_{|\pp R} = \mathrm{id}\}$. The ensuing reduced phase space $\Phi//\mathring\G$ leads to the ``edge-mode'' enhanced reduced  phase space of Maxwell-theory proposed e.g. in \cite{DonnellyFreidel}. Heuristically, edge modes can be identified with the pure-gauge part of $A$ at $\pp R$. A rigorous statement can be found in \cite{AldoNew}, where it was also proved that the resulting construction fails (in a precise sense) to be gauge invariant at the boundary.

Putting this issue aside, if we were to reduce $\Phi$ by bulk gauge transformations $\mathring\G$ only, the appropriate decomposition of $A$ would  be $A = \mathring A^\rad + \d \mathring \varsigma$ with $\mathring \varsigma$ defined by the boundary condition $ \mathring\varsigma_{|\pp R} = 0 $.
By duality, the corresponding decomposition $E=\mathring E_\rad + \pp \mathring\varphi$ would be defined by the boundary condition $\mathring\varphi_{|\pp R} = 0 $.
Interestingly, these boundary conditions for $\mathring\varsigma$ and $\mathring\varphi$ do not translate into natural boundary conditions for $\mathring A_i^\rad$ and $\mathring E^i_\rad$: in particular, notice that it is not true that $\mathring A_i^\rad{}_{|\pp R}$ is just $A_i{}_{|\pp R}$, nor that $s_i\mathring E^i_\rad{}_{|\pp R}$ is just $f$. Indeed, with this choice, the functional role of $f$ \emph{cannot} be clearly identified since both the Coulombic and the radiative components of $E$ contribute to it.
This state of affairs challenges the validity of naive (but fairly common) statements such that ``edge modes are conjugate to the flux $f$.''
\end{Rmk}

\paragraph{Physical interpretation of flux superselection}\label{sec:fluxSSS}
In this section we will attempt a physical interpretation of flux superselection.

The key to this interpretation is the observation that the symplectic structure over a bounded region $R$ describes only those degrees of freedom that evolve within the causal domain of $R$. That is, if $R$ is a $d$-ball $\bb B^d$, within the  $(d+1)$-dimensional causal diamond  $D(R)$ whose belt is $\pp R \cong \bb S^{d-1}$.
Observe that, to this spacetime region, the flux $f$ is associated in a completely invariant way:\footnote{Denoting with ${\cal F}$ the electromagnetic tensor, with a an under-arrow the pullback from the spacetime to $\pp R$, and with $\star$ the Hodge operator, the electric flux then reads: $f = \underleftarrow{\star \cal F}$.}
there is no need to select an arbitrary Lorentz frame, the $D(R)$ itself selects one for us.

This is as opposed to the case of a region $R$ that evolves in time into the spacetime cylinder $C\cong R\times \bb R$, where $\bb R$ stands for some time interval. The latter is e.g. the setup for a Casimir experiment, or a Faraday cage. But in this case, not only  the domain $C$ itself is not covariantly associated to $R$ (it is determined by the time evolution of the Faraday cage, say), and Lorentz frames along $\pp C \cong \pp R\times \bb R$ have to be continuously chosen to specify an $f$ at different times, but also the dynamics with $M$ is not determined unless one imposes specific boundary conditions at $\pp C$ \cite{Harlow_cov} (e.g. perfect conductor boundary conditions, e.g. \cite{BarnichCasimir:2019}). Focusing on the causal domain $D(R)$ therefore explains why, in the previous sections, we never had to mention any such boundary conditions: the dynamics within $D(R)$ is completely determined once $f$ is fixed. 

Therefore, we have argued, that the Maxwell dynamics itself is autonomous within $D(R)$ only at fixed $f$.
This way, it is consistent to associate distinct phase spaces, that is different superselection sectors, to different values of $R$.

There is, however, one remaining puzzle: if $f$ is fixed, does this mean that we are not allowed to ``create'' new charges within $R$? After all, adding a charge to $R$ means altering the electric flux through the boundary $\pp R$, doesn't it?
To see why this objection does not hold up to further scrutiny, one has to ask how much would the flux change at the ``creation'' of a new charge $q$ at $x\in R$. To compute the flux induced by the added charge one has to use the Gauss law, but, in the presence of boundaries, this elliptic equation can be uniquely solved only if a boundary condition is already given! 

(Of course, this boundary condition must be compatible with the integral Gauss law, which states that  the \emph{total} flux $\oint f$ is fixed by the total charge content of $R$, but is otherwise completely arbitrary. This means that the following argument, strictly speaking, hold for all but the zero-mode of $f$ over $\pp R$. Cf. Remark \ref{Rmk:Uniqueness}.)\footnote{In the non-Abelian theory, at the configuration $A$ there are as many integral Gauss laws as reducibility parameters for $A$ (a reducibility prarameter is a covariantly-constant Lie-algebra-valued function, so that $\delta_\chi A = \D \chi = 0$; it is the analogue of a Killing vector field in general relativity). Therefore, at a generic (i.e. non-reducible) non-Abelian configuration, there is \emph{no} integral Gauss law that needs to be satisfied: i.e. at a generic (i.e. non-reducible) non-Abelian configuration even the zero-mode of the electric flux is independent from the regional charge content. See also Footnote \ref{fnt:killing}. We refer to \cite[Sect.4]{GomesRiello-quasilocal} and references therein for a more thorough discussion of this point.}

In other words, the Coulombic field created by a charge $q$ at $x\in R$ can be computed throughout $R$  only in either of the following two circumstances: (\textit{1}) if we know the geometry of the whole Cauchy surface $\Sigma\supset R$, so that we can compute the Green's function of the boundary-less Poisson equation and thus solve the Gauss constraint throughout the entire $\Sigma$; or, (\textit{2}) if we know only the geometry of the subregion $R$, if we complement the Poisson equation by an appropriate boundary condition---such as we fix the flux $f$.
That is, the flux $f$ through $\pp R$ cannot be predicted from a viewpoint intrinsic to $ R$ not only because there might be charges on the outside of $R$ that affect it, but also because that flux depends on the (unknown) geometry of $\Sigma\setminus R$.
Conversely, there is no limit to the charges one can add in $R$ even at fixed flux $f$: the ensuing Coulomb potential can be simply computed through the Green's function associated to the flux $f$.

For a more thorough discussion of superselection in the context presented here, and comparisons with both algebraic QFT at asymptotic infinity and with the lattice, see \cite[Sec.7]{GomesRiello-quasilocal}.

\begin{Rmk}[Summary]
In this section we argued that the flux superselection structure that we have found by studying the symplectic geometry of Maxwell theory over $R$ is reflected in both the dynamics of Maxwell theory with the causal domain $D(R)$ and in the properties of the Green's functions necessary to solve the Gauss law intrinsically within $R$.
\end{Rmk}

\section{Gluing}\label{sec:Gluing}

\begin{figure}
\begin{center}
\includegraphics[width=.25\textwidth]{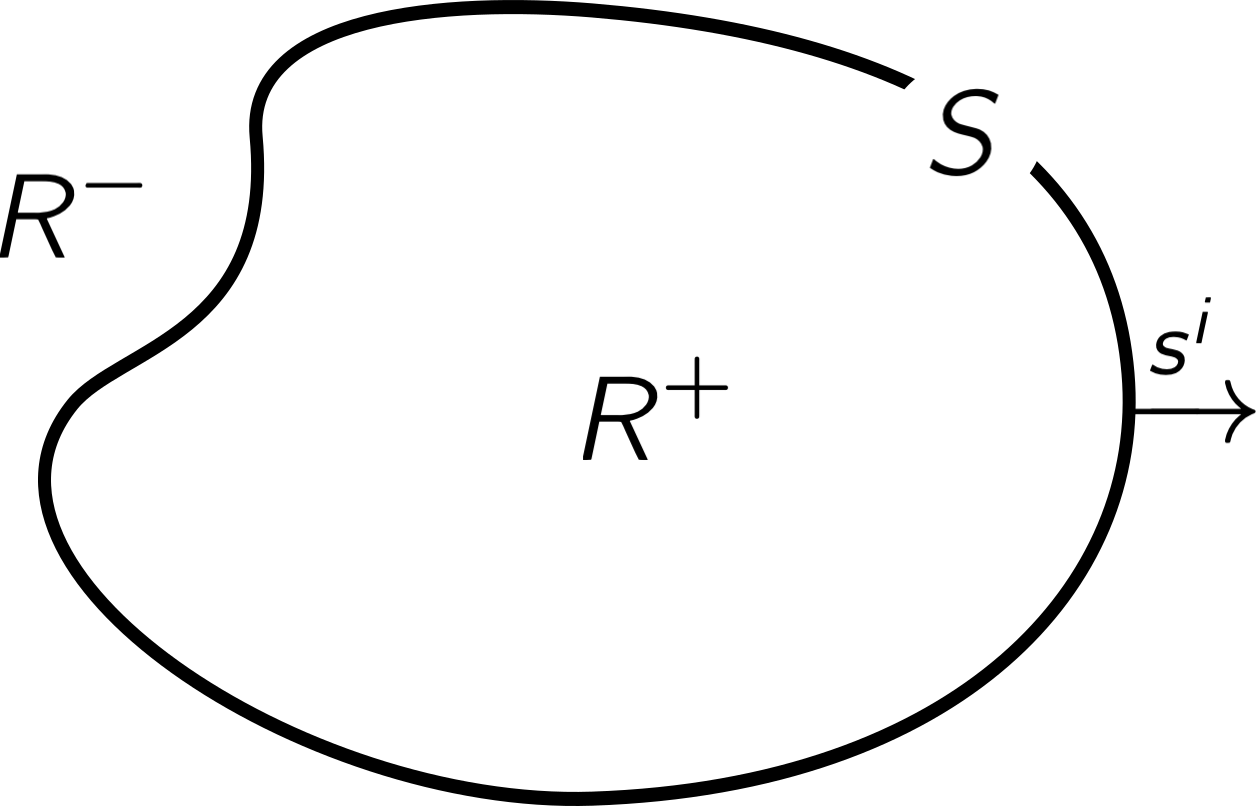}
\caption{The plane of the figure represents $\Sigma\cong \bb S^d$ (compactified at infinity). The two complementary subregions $R^\pm\subset\Sigma$ are glued along the interface $S$.} 
\label{fig:glue}
\end{center}
\end{figure}

\paragraph{The first and second gluing problems}
Consider now the Cauchy surface\footnote{To keep the notation streamlined, in this section we will denote covariant derivatives over $\Sigma$ with the symbol $\pp$. Similarly, we will omit volume forms from the integrals. We believe this abuse of notation will not create any confusion.}  $\Sigma \cong \bb S^{d}$, and split it in two ball-like regions $R^\pm \cong \bb B^d$ glued to each other along their common boundary, the interface $S = \pm\pp R^\pm \cong \bb S^{d-1}$. We will henceforth assume the boundary to be smooth, and that the unit normal $s^i$ to $S$ is taken outgoing from $R^+$, see Figure \ref{fig:glue}.

Consider now the Cauchy surface  $\Sigma \cong \bb R^{d}$, and split it into a ball-like regions $R^\pm$ glued to each other along their common boundary, the interface $S = \pm\pp R^\pm \cong \bb S^{d-1}$. We will henceforth assume the boundary to be smooth, and that the unit normal $s^i$ to $S$ is taken outgoing from $R^+$, see Figure \ref{fig:glue}.

In the previous section we analyzed the nature of the degrees of freedom of Maxwell theory in vacuum from a perspective intrinsic to each bounded region $R^\pm$.
We have found that these degrees of freedom organize in superselection sectors of fixed electric flux; in each sector it is not only the flux that is fixed but also, through the Gauss constraint, the entire Coulombic electric field throughout $R^\pm$.
In Coulomb gauge, the remaining degrees \emph{of freedom} that one can freely specify within a superselection sector are encoded in the radiative fields $(A^{\rad\pm}_i,E^i_{\rad\pm})$, characterized by the following properties
\be
\begin{cases}
\pp^i A_i^{\rad\pm} = 0 & \text{in }R^\pm\\
s^i A_i^{\rad\pm} = 0 & \text{at }S\\
\end{cases}
\qquad\text{and}\qquad
\begin{cases}
\pp_i E^i_{\rad\pm} = 0 & \text{in }R^\pm\\
s_i E^i_{\rad\pm} = 0 & \text{at }S\\
\end{cases}
\label{eq:rad}
\ee

We now ask, are these degrees of freedom sufficient to recover all the degrees of freedom present in the reduced phase space associated to the \emph{entire} Cauchy surface $\Sigma = R^+ \cup_S R^-$?  We call this \emph{the first gluing problem}.
 
Now, it is well-known that the gauge-invariant Hilbert space of lattice gauge theory does not factorize upon a partition of the lattice \cite{Polikarpov,Donnelly2008,DonnellyEntEnt:2011,Casini_gauge}. 
The classical, continuum, version of this statement is that the reduce phase space over $\Sigma$ does not factorize over the subregions $R^\pm$. denoted $\Omega^R_r$ the reduced symplectic structure over the region $R$ (without paying too much attention to the superselection structure), this can be schematically written as:
\be
\Omega^\Sigma_r \neq \Omega^+_r + \Omega^-_r.
\label{eq:nonfactoriz}
\ee
Because of this, it is legitimate to expect that new degrees of freedom must be included that do not belong to either reduced phase space. Identifying these extra degrees of freedom preventing the factorization of $\Omega_r^\Sigma$, constitutes what we call \emph{the second gluing problem}.

Now, it is often claimed that the first gluing problem is ambiguous, i.e. that extra degrees of freedom, e.g. the edge modes of Remark \ref{Rmk:edge}, are necessary to perform the gluing.\footnote{Another way to phrase this is that the gluing problem is ambiguous because the possibility of a “gauge slippage” at the interface cannot be avoided.}
And the main argument for this necessity is drawn from the evidence for the second gluing problem.

However, as we will prove shortly, the state of affairs is more subtle than that: although the \emph{second} gluing problem stands---i.e. new dof must indeed be included in the form of an extra term on the right-hand side of \eqref{eq:nonfactoriz} in order to reconstruct $\Omega^\Sigma_r$ from $\Omega_r^\pm$---the \emph{first} gluing problem is completely unambiguous---i.e. all degrees of freedom over $\Sigma$ \emph{can} be reconstructed from those over $R^+$ and $R^-$. 

In other words: in investigating the second gluing problem we will find that there \emph{is} a term which obstructs the factorization of the reduced symplectic structure over adjacent regions. And indeed, at a formal level, this term closely resembles an ``edge mode'' contribution. However, because by the first gluing problem all global dof \emph{are} actually reconstructable from the regional ones, that extra ``edge mode'' contribution \emph{cannot} involve new, functionally independent, degrees of freedom. That is, that ``edge mode'' contribution will be found to actually be a function(al) of the regional dof themselves.

This subtle state of affairs gives the title to this article, \emph{Edge modes without edge modes}, and clarifying it is the main goal of this section. 

\begin{Rmk}[Edge modes and gluing]
In this introduction we are \emph{not} claiming that the gluing procedure through edge modes is incorrect. We are claiming, rather, that is not ``minimal'' and in particular does not capture the most interesting features of the second gluing problem. Indeed, as explained in Remark \ref{Rmk:edge}, the edge-mode enhanced reduced phase space discussed in \cite{DonnellyFreidel} does not correspond to $\Phi^\pm//\G^\pm$ but rather to $\Phi^\pm//\mathring \G{}^\pm$. Heuristically, the latter space is the one obtained if one decides not to reduce with respect to boundary gauge transformations. Therefore, what the gluing procedure of \cite{DonnellyFreidel} implicitly advocates for is to first reduce with respect to bulk gauge transformations in either region, glue the resulting phase spaces, and finally reduce by the residual interface gauge transformations. This strategy closely parallels what one would do on the lattice, or for topological field theory, and there is nothing wrong with it (indeed, it makes one's life much easier in the presence of nontrivial topological cycles emerging upon gluing). But from the viability of this procedure it does \emph{not} logically follow that edge modes \emph{must} be included for gluing to be possible. In the next section, we will indeed prove that this is not the case: gluing \emph{can} be (surprisingly) successfully performed at the level of the fully reduced phase spaces.
\end{Rmk}

\begin{figure}
\begin{center}
\includegraphics[width=.9\textwidth]{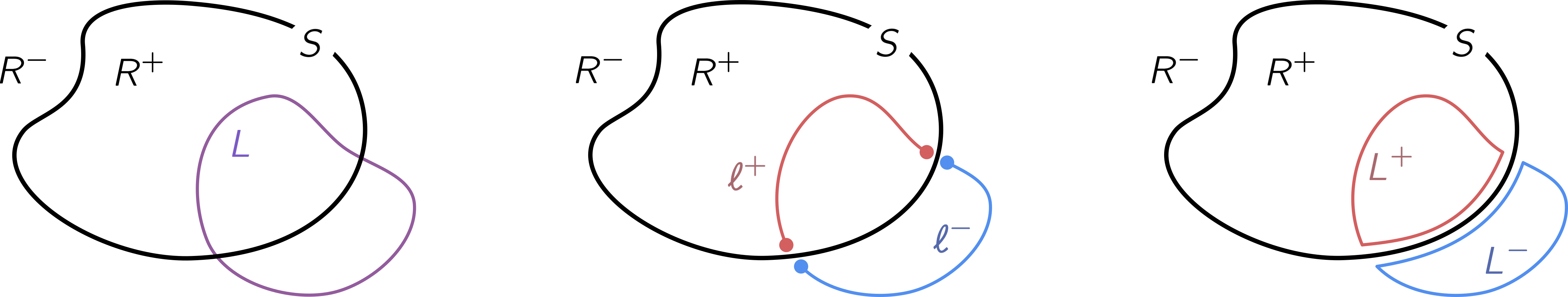}
\caption{A simple argument for the \emph{un}ambiguous nature of the first gluing problem. \emph{On the left}: It is depicted, in purple, a global gauge invariant observable: the Wilson loop $L[A] = \exp \oint_L A $. \emph{In the center}: The Wilson loop $L$ is split into two Wilson lines $\ell^\pm[A] = \exp\int_{\ell^\pm} A$ (in red and blue), each line supported in one of the subregions $R^\pm$; the global observable can be reconstructed from the regional ones, $L = \ell^+\times \ell^-$, but the latter fail to be gauge invariant---unless extra degrees of freedom, the edge modes, are introduced at its ends (represented by dots). \emph{On the right}: However, by appropriately closing the Wilson lines $\ell^\pm$ into Wilson loops $L^\pm$, not only there is manifestly no need to introduce extra degrees of freedom to ensure gauge invariance, but it is also still possible to reconstruct the global observable from the regional ones: $L = L^+ \times L^-$. (Notice that, on the lattice, a tension arises between the central and right-most pictures, due to the fact that there one must decide wether to split the lattice across links or faces---cf. \cite{Casini_gauge}.)} 
\label{fig:loops}
\end{center}
\end{figure}

\paragraph{The first gluing problem}\label{sec:1stgluing}
Being Abelian, in Maxwell theory there is a very simple argument hinting that the fist gluing problem should be completely \emph{un}ambiguous. 
The argument focuses on the magnetic degrees of freedom, and shows that there is no ``boundary gauge'' ambiguity that arises when attempting to glue gauge-invariant degrees of freedom across the interface $S$.
The argument is most easily expressed in pictures, and we therefore refer to Figure \ref{fig:loops} and its caption.

Despite the simplicity of the Wilson-line argument, it is still important to provide a formal solution to the first gluing problem. Indeed this solution will not only helping solve the second gluing problem explicitly, but, when applied to the electric degrees of freedom, it will also help us shedding new light on flux superselection.

In what follows we will denote by $h_\pm$ ``regional radiative'' 1-tensors, i.e. 1-tensors that satisfy equations \eqref{eq:rad} over $R^\pm$.
Similarly, we will denote by $H$ ``global radiative'' 1-tensors, i.e. 1-tensors over $\Sigma$ which are divergence-free:
\be
\pp_i H^i = 0 \quad \text{in }\Sigma
\label{eq:H}
\ee
(since $\pp\Sigma=\emptyset$, $H$ does not need to satisfy any boundary condition).
We will also denote by $\Theta_\pm$ the characteristic functions of $R^\pm$: e.g. $\Theta_+(x) = 1$  if $x\in R^+$ and $\Theta_+(x) = 0$ otherwise.

For simplicity of exposition, let us start from the ``inverse'' of the gluing problem. 
Suppose to be given a global radiative $H$; its restrictions $ H\Theta_\pm$ to either $R^\pm$ can be themselves decomposed into \emph{regional} radiative and Coulombic components $h_\pm $ and $\lambda_\pm$:
\be
H = (h_+  + \pp \lambda_+)\Theta_+ + (h_-  + \pp \lambda_-)\Theta_-,
\label{eq:Hhxi}
\ee
where
\be
\begin{cases}
 \pp_i h^i_\pm = 0 & \text{in } R^\pm\\
 s_ih^i_\pm = 0& \text{at }S
\end{cases}
\quad\text{and}\quad
\begin{cases}
\Delta \lambda_\pm =  \pp_i H^i =0& \text{in } R^\pm\\
s^i \pp_i \lambda_\pm = \pm s_iH^i & \text{at }S
\end{cases}
\label{eq:gluing1}
\ee  
Clearly, given $h_\pm\neq H_\pm$, a fact that can be summarized as ``projection (onto the radiative piece) does not commute with restriction (to $R^\pm$).'' Nonetheless, given $H$, the regional radiative components $h_\pm$ can be explicitly computed by solving the EBVP on the right,\footnote{The minus $\pm$ sign in the boundary condition to the Poisson EBVP is due to the fact that $s^i$ is by convention the outgoing normal to $R^+$.}
which gives us $\lambda_\pm = \lambda_\pm(H)$ and thus $h_\pm = h_\pm(H) = H\Theta_\pm - \pp \lambda_\pm(H)$.
Therefore, the ``inverse'' of the first gluing problem is immediately solved.

The first gluing problem asks whether equations (\ref{eq:H}-\ref{eq:gluing1}) can be solved not for $h_\pm$ as a function of $H$, but for  $H$ as a function of $h_\pm$. In other words, it asks whether the global radiative $H$ can be uniquely reconstructed from the regional radiatives $h_\pm$.
Remarkably, the answer is affirmative.\footnote{We are here not interested in the existence part of the proof, just in its uniqueness. We will assume in the following that $H$ is at least $C^1$ in $\Sigma$.}

A proof of this statement can be found in \cite[Sec.6]{GomesRiello-quasilocal}, here we will simply report the result and give a rough sketch of the proof. 
The result is expressed in terms of the following system of equation that uniquely fixes the $\lambda_\pm = \lambda_\pm(h_+,h_-)$ that allow to invert \eqref{eq:Hhxi} for $H = H(h_+,h_-)$:
\be
\begin{cases}
\Delta \lambda_\pm = 0& \text{in } R^\pm\\
s^i \pp_i \lambda_\pm = \Pi & \text{at }S
\end{cases}
\label{eq:rec1}
\ee
where\footnote{Of course this $\Pi \in C(S)$ has nothing to do with the Poisson bivector $\Pi^{IJ} \in \mathfrak{X}^{\wedge 2}(\Phi)$ discussed in Section \ref{sec:thingssymp}.}
\be
\Pi = - (\mathcal R_+^{-1} + \mathcal R_-^{-1} )^{-1} \mu
\qquad\text{and}\qquad
\Delta_S\mu = \div_S(h_+ - h_-)_{|S}.
\label{eq:Pi}
\ee
These equations involve many unexplained symbols, and therefore require a few explanatory comments.
\begin{enumerate}[label=\roman*.,font=\itshape]
\item First of all, just as in \eqref{eq:gluing1}, due to the divergence-free property of both $h_\pm$ and $H$, the $\lambda_\pm$'s must satisfy a Laplace equation.
Moreover, due to the fact that at $S$ both $s_i h^i_\pm = 0$, the continuity of $H$ demands that the two regional Laplace equations share the same Neumann boundary condition (up to a sign).
We denote this boundary condition by $\Pi$---the next two points are aimed at fixing $\Pi$ in terms of $h_\pm$.
Notice that, as usual, once $\Pi$ is fixed, the Neumann EBVP fixes the $\lambda_\pm$'s up to a constant---which is anyway irrelevant for the sake of reconstructing $H$. \label{comment1}

\item Using the fact that at $S$ both $s_i h^i_\pm = 0$, we find that both the $h_\pm$ are parallel to $S$ and therefore the \emph{radiative mismatch} $(h_+ - h_-)_{|S}$ is a 1-tensor intrinsic to $S$. Similarly introducing the \emph{Coulombic mismatch}
\be
\mu := -(\lambda_+ - \lambda-){}_{|S},
\ee 
one sees that continuity of $H$ across the interface, leads also to the above \emph{relationship between the radiative and Coulombic mismatch} (the second equation in \eqref{eq:Pi}). Notice that this relationship is expressed in terms of yet another elliptic Poisson equation, this time \emph{intrinsic} to the interface $S$ (there, $\Delta_S$ and $\div_S$ are the intrinsic Laplacian and divergence operators to $S$).

\item Finally, one needs to convert the Coulombic \emph{mismatch} into a boundary condition for either $\lambda_\pm$. This is possible because we already know that they both satisfy the same Neumann boundary condition $\Pi$. This is what allows us to fix $\Pi$ as a function of $\mu$. The operator that allows to perform the conversion is a combination of so-called Dirichlet-to-Neumann pseudo-differential operators $\mathcal R_\pm$. These operators act on functions defined on the boundary of a region, and map Dirichlet boundary condition onto Neumann boundary conditions, so that
\be
V
\qquad\text{satisfies}\qquad
\begin{cases}
\Delta V = 0& \text{in } R^\pm\\
V = u & \text{at }S
\end{cases}
\qquad\text{iff}\qquad
\begin{cases}
\Delta V = 0& \text{in } R^\pm\\
s^i\pp_iV = \mathcal R_\pm(u) & \text{at }S
\end{cases}
\ee
One can check that the operators $\mathcal R_\pm$ are linear, positive definite, and self-adjoint with respect to the natural $L^2$ Riemann-measure intrinsic to $S$. These facts ensure that the combination $(\mathcal R_+^{-1} + \mathcal R_-^{-1} )$ defines an invertible operator on the functions over $S$.  This concludes our explanation of the reconstruction formulas \eqref{eq:rec1} and \eqref{eq:Pi}.
\end{enumerate}

\begin{Rmk}[Summary]
To summarize, we have found that the two regional radiatives $h_\pm$ contain all the information necessary for the complete reconstruction of the global radiative $H$.
This reconstruction involves the inversion of several differential and pseudo-differential operator, and is therefore quite non-local. At its core, however, it says that the difference between $H$ and $h_\pm$ can be fully reconstructed from the \emph{radiative mismatch} $(h_+ - h_-)_{|S}$ at the interface $S$. 
Notice how this mismatch depends on \emph{both} $h_+$ and $h_-$, \emph{both} of which are thus required to reconstruct $H$ in either region. 
\end{Rmk}

The next two remarks go at the core of our treatment of gluing.

\begin{Rmk}[Gluing gauge potentials]\label{Rmk:gluingA}
We now interpret the above results in terms of the gluing of the gauge potential $A$, i.e. upon the identification $H\leadsto A^\rad$ and $h_\pm = A^{\rad\pm}$. In this case the $\lambda^\pm\leadsto\varsigma_\pm$ stand for ``pure gauge'' adjustments that allow one to translate between the global Coulomb gauge satisfied by $A^\rad$ to the regional Coulomb gauges satisfied by $A^{\rad\pm}$. 
However, these are \emph{not quite} gauge transformations as we defined them in the previous sections: these $\varsigma_\pm$'s are indeed functionals of the radiative gauge fields! In other words the $\varsigma_\pm$'s are functionals of the configuration in the \emph{reduced} phase spaces $A^{\rad\pm}\in\Phi^\pm\slashf\G^\pm$. As such, these functionals represent \emph{physical} quantities. 

We postpone their physical identification until the end of the next remark.
\end{Rmk}

\begin{Rmk}[Gluing electric fields]\label{Rmk:gluingE}
We now interpret the above results in terms of the gluing of the electric fields $E$, i.e. upon the identification $H\leadsto E_\rad$ and $E_\pm = E_{\rad\pm}$. Since we are working on-shell of the Gauss constraint, in this case the $\lambda_\pm$'s stand for ``pure Coulombic'' adjustments that allow one to translate between the global Coulomb potential, $\varphi = 0$ in vacuum, and the regional ones $\varphi_\pm = \lambda_\pm(E_{\rad+},E_{\rad-})$.
But as we repeatedly emphasized, e.g. in section \ref{sec:fluxSSS}, $\varphi_\pm$ is completely fixed by the electric flux $f$. And indeed, a comparison between \eqref{eq:EBVPvarphi} and the reconstruction equations \eqref{eq:rec1} immediately unveils the correspondence $\Pi \leadsto f$.
Notice what happened here: whereas the flux $f$ is functionally independent of either $E_{\rad+}$ \emph{or} $E_{\rad-}$, it is indeed a functional of the electric \emph{radiative mismatch} $(E_{\rad+}-E_{\rad-})_{|S}$! 
That is, knowledge of the electric radiative modes in \emph{both} region allows 	the full reconstruction of the flux $f$ through the common interface $S$ and hence of the Coulombic fields \emph{throughout} both regions $R^\pm$.

Finally, since in Coulomb gauge $E_\rad = \dot A^\rad$, it is clear that the quantity $\Pi_A := s^i\pp_i\varsigma^{\rad\pm}$---computed from the radiative mismatch $(A^{\rad+}-A^{\rad-})_{|S}$ as in \eqref{eq:Pi}---is the time-antiderivative of the electric flux, that is $f = \dot\Pi_A$. This confirms the physical, gauge invariant, nature of the adjustments $\varsigma_\pm$.
\end{Rmk}

\begin{Rmk}[Superselection, revisited]
At the light of the previous remark we see how the regional Coulombic components of the electric field, which used to be superselected ``before gluing,''  turn into dynamical variables upon gluing!
This is perfectly consistent because the dynamics at the interface $S$ now lies within the causal domain of the unions of the two regions $S \subset D(\Sigma = R^+\cup R^-)$---thus avoiding all contradiction with the discussion of section \ref{sec:fluxSSS}.
One could summarize this state of affairs by stating that \emph{flux superselection is a consequence of ``tracing over'' the degrees of freedom in the exterior of a region $R$}. 
This viewpoint, consistent with the lattice perspective \cite{DonnellyEntEnt:2011,Casini_gauge,Delcamp:2016eya}, also suggests a prominent role for the Coulombic electric field in the computation of the entanglement entropy. In this regard, it would be interesting to reproduce the results of \cite{Donnelly:2014fua,AronWill} through a real-time entropy computation, rather via replica-trick, since one expects the real-time computation to benefit from the same radiative/Coulombic decompositions investigated here.
\end{Rmk}

\begin{Rmk}[Entanglement entropy and the replica trick]
Regarding entanglement entropy, the authors of  \cite{Donnelly:2014fua,AronWill} come to a very similar conclusion to ours regarding the nature of the ``edge mode contribution'' to entanglement entropy---which they show is related to the Kabat contact term. Indeed, despite appearances, what they call the ``edge mode contribution'' in  \cite{Donnelly:2014fua,AronWill}  does \emph{not} quite come from would-be gauge boundary dof (which is how the phrase ``edge modes'' seems to be most commonly used these days \cite{DonnellyFreidel}, and how we used it in this article too).

In fact, in  \cite{Donnelly:2014fua,AronWill}, ``edge modes'' are rather defined as: ``\emph{the unique static classical solution of the form $E = \nabla\phi$ with the boundary condition $\nabla_\perp\phi = E_\perp$}'' [cit.]. This is extremely close to what we are proposing here: that the presence of a superselected Coulombic electric field is what characterizes Maxwell theory in the presence of boundaries, and by extension what we expect distinguishes its entanglement entropy from that of, say, a scalar field theory.

However, there is a subtle difference between our and their proposals: whereas we are presently concerned with the electromagnetic fields \emph{over $R$}, the authors of  \cite{Donnelly:2014fua,AronWill} compute the entanglement entropy through a replica trick and therefore are concerned with the electromagnetic field over an Euclidianized spacetime with conical singularity at $\pp R\times \bb S^1$. In other words, the authors of \cite{Donnelly:2014fua,AronWill} defined their ``Coulombic modes'' over an auxiliary space which possesses an extra Euclidean $\bb S^1$ direction.

 This said, in their explicit computation, the extra $\bb S^1$ direction ends up playing a spectator role, in the sense that it only contributes an overall global factor of $\beta$ (inverse temperature) that multiplies what we would call here the total Coulombic energy\footnote{The notion of energy is tied to a notion of time evolution. The relevant notion here is that of Rindler time. Since the Rindler lapse vanishes at the Rindler horizon, a regularization procedure is necessary.} in a given super-selection sector. In sum, the edge-mode contribution to the entanglement entropy in  \cite{Donnelly:2014fua,AronWill} (heuristically) corresponds to the thermal canonical entropy associated to each superselection sector computed at the geometric temperature $\beta=2\pi$. 

Therefore, the framework proposed in this article is closely related to the discussion of the Kabat contact term in  \cite{Donnelly:2014fua,AronWill}. Hence the interest of recovering their results in a physically more transparent way through a ``real time'' computation (i.e. without replica trick) and without ever referring to the lattice to justify the need of superselection or the form of the path-integral measure. We believe the formalism presented in this article should provide the necessary tools to achieve these results.
\end{Rmk}

\paragraph{The second gluing problem: or, edge modes without edge modes}\label{sec:EMWEM}
The second gluing problem consists in identifying precisely which degrees of freedom present in $\Omega^\Sigma_r$ lack in the sum of  $\Omega^\pm_r$. 
The Remarks \ref{Rmk:gluingA} and \ref{Rmk:gluingE} above already answer to this question, identifying the missing degrees of freedom in the quantities $\Pi_A$ and $\Pi_E = f$---now understood as functionals of the radiative mismatches  $(A^{\rad+}-A^{\rad-})_{|S}$ and $(E_{\rad+}-E_{\rad-})_{|S}$.

In this brief section, we will simply derive how these degrees of freedom enter the symplectic structure $\Omega^\Sigma_r$ when expressed in terms of $(A^{\rad\pm},E_{\rad\pm})$.

We start from the on-shell symplectic structure over $\Sigma$ \eqref{eq:OmegaRad}, that is
\be
\pi^*\Omega_r =  \int  \dd E^i_{\rad} \curlywedge \dd A_i^{\rad}.
\ee
We then decompose the fields $(E^i_{\rad} , A_i^{\rad})$ over their regional radiative and Coulombic components, as in \eqref{eq:Hhxi}, thus obtaining
\begin{align}
\pi^*\Omega_r &= \sum_\pm \int_\pm \dd E^i_{\rad\pm} \curlywedge \dd A_i^{\rad\pm}  + \oint  \dd f \curlywedge \dd \mu_A
\qquad\text{where}\qquad
\mu_A := (\varsigma_+ - \varsigma_-)_{|S}
\label{eq:gluing2}
\end{align}

Comparing with equations \eqref{eq:iOmegaf} and \eqref{eq:Omegafr}, we see that the first term on the right-hand side of the this equation coincides with the radiative part of $\pi^*\Omega^f_r$. However, they are \emph{not the same} since $\pi^*\Omega^f_r$ implicitly assumes that $f$, and thus $\varphi_\pm(f)$, are fixed, whereas in \eqref{eq:gluing2} the flux $f$ is completely ``unfrozen'' and free to vary---\emph{with a catch}: both the flux $f$ and the Coulombic mismatch $\mu_A$ must be understood in \eqref{eq:gluing2} as functionals of the (mismatch between the) radiative degrees of freedom $(E^i_{\rad\pm} , A_i^{\rad\pm})$, and \emph{not} as extra degrees of freedom independent of $(E^i_{\rad\pm} , A_i^{\rad\pm})$.

For these reasons, equation \eqref{eq:gluing2} should be more explicitly be written as
\be
\pi^*\Omega_r = \sum_\pm \int_\pm \dd E^i_{\rad\pm} \curlywedge \dd A_i^{\rad\pm} 
+ \oint     \big(\mathcal R_+^{-1} + \mathcal R_{-}^{-1}\big)^{-1}  \frac{\div_S[\dd E_\rad]^\pm}{\Delta_S^2} \curlywedge  \frac{\div_S[\dd A_\rad]^\pm}{\Delta_S^2},
\ee
where we used the short-hand notation $[h]^\pm:=(h^+ - h^-){}_{|S}$.\footnote{Notice that the differential $\dd$ ``go through'' the integral and (pseudo)differential operators $\Delta_S^{-1}$, $\div_S$, $\mathcal R_\pm$, because they are all field-independent, i.e. do not depend on the Maxwell configuratiion $(A,E)$. This is the case in Abelian theory only. See Section \ref{sec:YM} and \cite[Sec.6]{GomesRiello-quasilocal} for details on the non-Abelian case.}

\begin{Rmk}[Edge modes without edge modes]
\emph{Naively}, equation \eqref{eq:gluing2} could be  misread as stating the necessity of a new boundary degree of freedom $\mu_A$ related to the pure-gauge part $\varsigma$ of $A$ and conjugate to $f$.
However, matters are more subtle.
From a \emph{global} perspective, $f$ and $\mu_A$ are to be understood as functionals of the regional radiatives, and not as independent degrees of freedom.
Conversely, from a regional perspective---i.e. after having ``traced over'' the regional degrees of freedom over, say, $R^+$---the flux $f$ becomes superselected and the mismatch $\mu_A$ ceases to be relevant (one could say it ``becomes pure gauge''). This is why equation \eqref{eq:gluing2}  can be understood as the ``\emph{edge modes without edge modes}'' formula that gives the title to this article.
\end{Rmk}

\section{Maxwell theory with matter}\label{sec:MaxMatt}
Introducing matter fields in the analyses of the previous sections is relatively straightforward, and we refer to \cite{GomesRiello-quasilocal} for details. 
Here, we will limit ourselves to stressing three aspects which make the charged case stand apart.

\paragraph{The Coulombic electric field}
In the absence of matter, the Gauss constraint entirely determines the Coulomb potential $\varphi$ in $R$ in terms of the electric flux $f$. Hence, superselection of $f$ means superselection of $\varphi(f)$ throughout $R$.

In the presence of matter, however, the Gauss constraint determines the Coulomb potential $\varphi$ in $R$ in terms of the electric flux $f$ \emph{and the regional charge density $\rho$}.
Hence, since $\rho$ is a gauge invariant and local quantity in $R$ and it is therefore part of the regional reduced phase space, $\varphi=\varphi(f,\rho)$ is no longer superselected, in particular $\varphi$ does not Poisson-commute with the charged matter fields. 

Nonetheless, the superselection of $f$ is untouched.

\paragraph{Dressed matter fields}\label{sec:dressing}
The charged matter-field analogue of the radiative field $A^\rad$ is the bounded-region Dirac-dressed electron $\hat \psi$ \cite{Dirac:1955uv} (see also\cite{GomesRiello2018}, \cite[Sect.9]{GomesHopfRiello}, and in particular \cite[Sec.5]{GomesRiello-quasilocal}). 
Schematically, denoting $\varsigma(A)$ the solution to \eqref{eq:dressing}, one has: $A^\rad = A - \d \varsigma(A)$ and\footnote{Recall that in our conventions, $A$ and $\sigma$ are already imaginary-valued, since we took $\Lie(\mathrm U(1)) \cong \mathrm i \bb R$. Here, $q$ is the charge of the particle described by the field $\psi$.}
\be
\hat \psi = e^{q \varsigma(A)} \psi.
\label{eq:dressedpsi}
\ee
If $R \cong \bb R^3$, and all fields satisfy fast fall-off conditions at infinity, it is easy to see that $\hat \psi$ is indeed Dirac's dressed electron. 

The particular field-dependence of $\varsigma(A)$ has three main consequences:
\begin{enumerate}[label=\roman*.,font=\itshape]
\item the Coulombic field $\varepsilon^i_\mathrm{Coul}(x|y)$ at $x$ associated to a charged electron $\hat\psi$ at $y$---which can be  computed through the Poisson bracket $\{E^i(x) ,\hat\psi(y)\} = - \varepsilon^i_\mathrm{Coul}(x|y) \hat \psi(y)$---carries a flux that affects only the zero-mode of $f$, consistently with the discussion of Section \ref{sec:fluxSSS};
\item it makes the (composite) field $\hat \psi$ gauge invariant, i.e. it makes $\hat \psi$ a good coordinate on the reduced phase space;
\item it makes $\hat\psi$ charged (of charge $q$) under the action of ``global gauge transformations'', i.e. under the action of the charge group $\mathrm{U}(1)$.
\end{enumerate}
Notice that the third point is not in contradiction with the second one, because only ``global gauge transformations'' are in the kernel of \eqref{eq:dressing}. 
For more on this cf. Remark \ref{Rmk:Uniqueness} and Footnote \ref{fnt:killing}.

\paragraph{Gluing with matter}
For the same reason that $\hat \psi$ is gauge-invariant but charged under the (global) charge group $\mathrm{U}(1)$, gluing in the presence of matter fields can be ambiguous. 
Indeed, the reconstruction theorem uniquely determines $\pp\lambda_\pm$, whereas $\lambda_\pm$ are determined only up to a spatial constant (see comment \textit{\ref{comment1}} below \eqref{eq:Pi}).
Hence, whereas $A^\rad$ can be uniquely reconstructed, $\hat \psi$ can suffer ambiguities in the reconstruction related to regional phase-shifts of its charges. This situation is best illustrated with an example.

\begin{figure}
\begin{center}
\includegraphics[width=.3\textwidth]{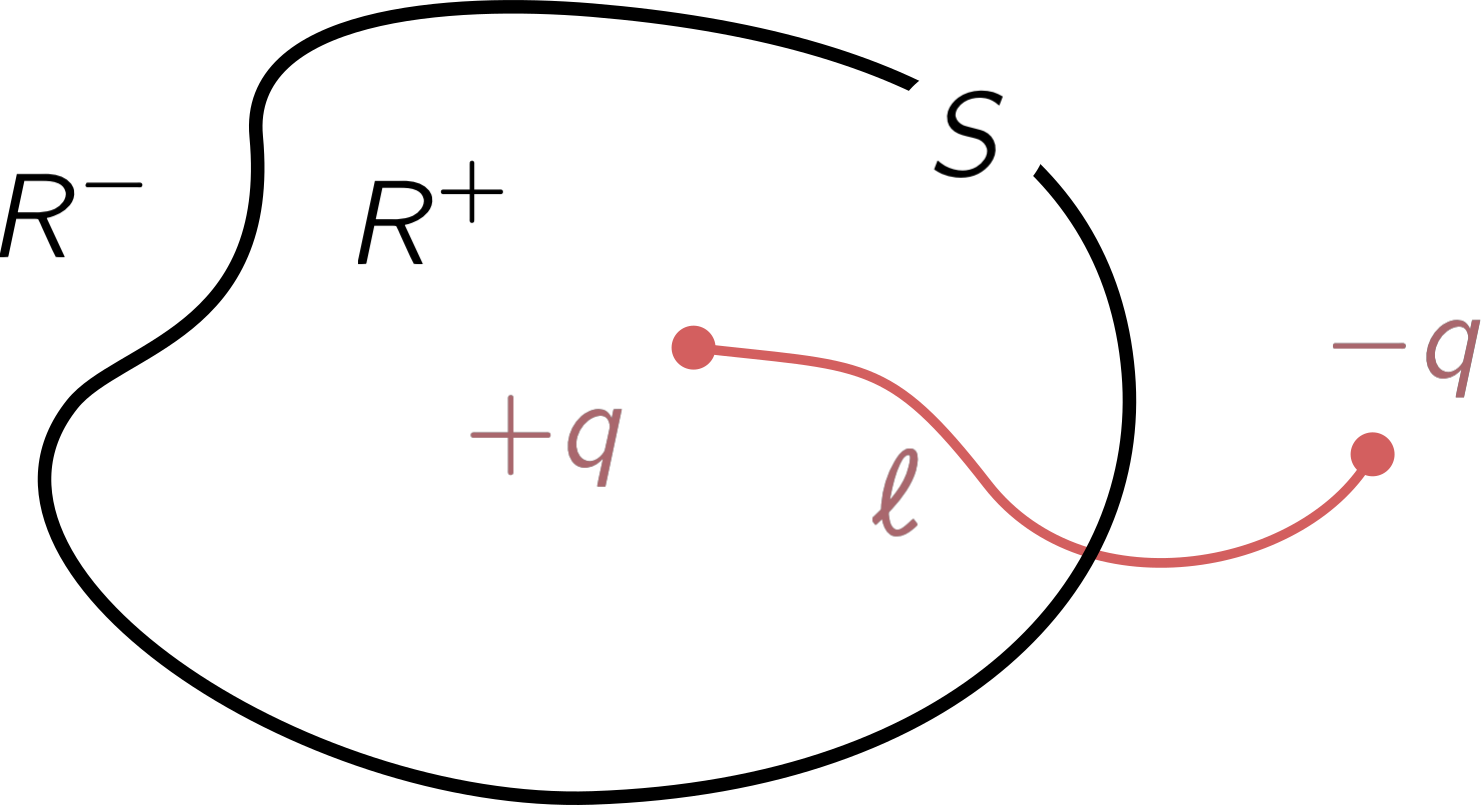}
\caption{Two complementary regions $R^\pm$ containing two opposite point-like charges connected by a Wilson line $\ell$. The relative phase between the two charges is not fixed by the gluing formulas of Section \ref{sec:1stgluing}. This phase, however, has physical relevance since it can be detected through interference experiments or through gauge-invariant Wilson-line observables.} 
\label{fig:charges}
\end{center}
\end{figure}

Consider the gluing of two regions with two opposite point-like charges in their interiors, as depicted in Figure \ref{fig:charges}.
The gluing formulas of section \ref{sec:1stgluing}, determine the quantities $\lambda_\pm\leadsto\varsigma_\pm(A)$ only up to a spatial constant $c_\pm$.
Therefore, although $A^\rad$ is unambiguously reconstructed from $A^{\rad\pm}$---this is because these reconstruction formulas involve the gradient of $\varsigma_\pm(A)$---the charges' dressed fields $\hat \psi_\pm$ determine the total one only up to a spatially-constant phase shift:
\be
\hat \psi = e^{ q (\varsigma_+(A) + \mathrm i c_+)}  \hat\psi_+  + e^{ - q (\varsigma_-(A) + \mathrm i c_-)}  \hat\psi_- .
\ee
Neglecting an irrelevant global phase shift, the relative phase shift, equal to $\phi = \mathrm i q (c_+ + c_-)$, is physical since it can be detected in an interference experiments like 't Hooft's beam splitter \cite{thooft1980beam} or by computing the gauge-invariant Wilson line observable
\be
W_\ell = \bar \psi_- e^{\int_\ell A} \psi_+ \leadsto e^{\mathrm i \phi} W_\ell
\ee
This provides further evidence of the different status \emph{global} gauge symmetries possess with respect to ``local'' ones (see Footnote \ref{fnt:killing} for a precise distinction between the two).
 
This situation is similar to Galileo's ship thought experiments: although the absolute positions of the ship ($\sim$ the phase of $\hat \psi_+$) and that of the pier ($\sim$ the phase of $\hat\psi_-$) have no physical bearing as long as experiments are conducted within those reference systems, the relative position between the two (=the relative phase $\phi$) has physical relevance when experiments involving both reference systems are performed, such as throwing a rope to anchor the ship to the pier. In this analogy, ``local'' gauge symmetries correspond to diffeomorphisms of the flat and translation invariant Euclidean space in which the ship and the pier are located. For more on this, and the relevance of this example for the discussion of the direct empirical significance (DES) of gauge symmetries,\footnote{For more on DES, and DES of gauge symmetries in particular, see \cite{Gomes:2019otw} and references therein. See also both the introduction and the last section of \cite{strocchi2015symmetries}.} see \cite[Sec.6.4]{GomesRiello-quasilocal}.

\section{Generalization to non-Abelian Yang--Mills theory\label{sec:YM}}
If inclusion of matter presented hardly any technical hurdle, generalization of the previous analyses to non-Abelian Yang--Mills theory requires a major sharpening of the mathematical tools employed. 
This is where the ``functional connection'' subject of most of our previous work comes into play \cite{GomesRiello2016,GomesRiello2018,GomesHopfRiello,GomesRiello-quasilocal,AldoNew}.
As it turns out, this functional connection is also the key to relaxing our reliance on any particular gauge fixing, and proving that our results are fully invariant.

\paragraph{Notation} 
In the non-Abelian case we will adopt wherever possible notations that parallel those already employed in the Abelian case. 
We will henceforth assume $G= \SU(N)$, so for example: $\A = \Omega^1(R,\mathfrak g)\ni A$, $\Phi = \T^*\A\ni(A,E)$, $\G = \Omega^0(R,G)\ni g$, $\Lie(\G) = \Omega^0(R,\mathfrak g) \ni \xi$, and so on.
We also denote the covariant derivative $\d + A$ as $\D$, leaving understood the representation in which it acts, e.g.
\be
\D \xi = \d \xi + [A, \xi].
\ee
Finally, the formulation of Yang--Mills theory relies on the existence of a preferred non-degenerate $\Ad$-invariant bilinear form $\tr(\cdot\,\cdot)$ over $\mathfrak g$ (the Killing form). 
In particular, this bilinear form allows us to identify---as usual!---$E\in \T_A^*\A$ with a vector field valued in $\frak g$ (denoted with the same symbol), so that the canonical 1- and 2-forms can be written as:
\be
\theta = \int \tr(E^i \dd A_i) \qquad\text{and}\qquad\Omega = \int \tr(\dd E^i\curlywedge \dd A_i).
\ee
Finally, in the non-Abelian theory, the Gauss constraints reads
\be
\GC := \D_iE^i \equiv \pp_i E^i + [A_i, E^i].
\ee

\paragraph{Covariant superselection sectors}
The first most obvious, and most important, feature that makes the non-Abelian case stand out is that the electric flux $f$ is not gauge invariant (it transforms in the adjoint representation) and therefore cannot be fixed in a superselection sector without breaking gauge symmetry at the boundary.
This will have deep consequence for the theory of non-Abelian superselection sectors which must be now identified with \emph{conjugacy classes} 
\be
[f] := \{ f' : \exists g\in\G \text{ such that } f' = \Ad_{g|\pp R} f \}.
\label{eq:classf}
\ee
We will denote the space of on-shell configurations  with flux $f$ belonging to a given conjugacy class $[f_o]$ by $\Phi^{[f_o]}_o$, and call its quotient by gauge transformations a \emph{covariant superselection sectors},
\be
\Phi//_{\hspace{-2pt}[f]\hspace{2pt}} \G := \Phi_o^{[f_o]}/\G.
\ee

Now, since the pullback of $\dd f$ to $\Phi^{[f_o]}_o$ clearly fails to vanish, the kernel of $\iota_{[f_o]}^*\Omega$ will fail to include all gauge transformations. 
In \cite{AldoNew}, we showed that this difficulty can be resolved by ``completing'' the 2-form $\iota_{[f_o]}^*\Omega$ by adding to it the canonically-given (infinite dimensional analogue of the) KKS symplectic 2-form \cite[Ch.14]{Marsden_1999}, 
\be
\omega_\text{KKS}^{[f_o]}\in\Omega^2([f_o]).
\ee
(That this works, is a fact which is not obvious in the present context). Notice that it is \emph{the fully canonical\hspace{1pt}\footnote{In the sense of intrinsic, independent of any external input.}\hspace{-1pt} nature of this completion which makes it viable.} 

To see this recall from Section \ref{sec:SymplFol} that the dual of a Lie algebra $\mathfrak g^*$ admits a canonical symplectic foliation by its coadjoint orbits. Here, any $f_o$ can be understood as an element of $\Lie(\G_{|\pp R})=\Omega^0(\pp R, \mathfrak g)$ and $[f_o]$ as an adjoint orbit\footnote{Here $\G_{|\pp R}$ is the restriction of the full (bulk) gauge group $\G$ to the boundary (or, more precisely, $\G_{|\pp R} := \G/ \mathring\G$), and \emph{not} the group of ``boundary gauge transformations.'' In particular, this means that $\G_{|\pp R}$ contains \emph{no} ``large" (boundary) gauge transformations which by definition fail to be connected to the identity. Consequently, $[f_o]$ is always connected. Cf. the definition \eqref{eq:classf} and, for more details, \cite{AldoNew} and \cite{GomesRiello:theta}.\label{fnt:LieppR}} therein. Therefore, up to the dualization, this is just an infinite dimensional lift of the previous finite-dimensional discussion. Concerning the dualization, the bilinear form $\tr(\cdot\,\cdot)$ can be used here to identify adjoint and coadjoint orbits.
Hence, the 2-form $\omega_\text{KKS}^{[f_o]}\in\Omega^2([f_o])$ is canonically given on each (co)adjoint orbit $[f_o]$, that is, it is naturally associated to $[f_o]$ and does not require any extra input from our part.  
For an explicit expression, see \cite{AldoNew}.

In more detail, what we showed in \cite{AldoNew} is that the reduced space $\Phi//\G$ \emph{admits a (natural!) symplectic foliation by the covariant superselection sectors}
\be
\big(\Phi//_{\hspace{-2pt}[f]\hspace{2pt}} \G, \Omega_r^{[f]}\big)_{[f]},
\ee
where the symplectic 2-form $\Omega_r^{[f]}$ is obtained by projecting the following 2-form over $\Phi_o^{[f]}$:
\be
\pi^*\Omega^{[f]}_r = \iota_{[f]}^* \Omega + \omega_\text{KKS}^{[f]}.
\label{eq:KKScomplete}
\ee

\paragraph{Functional connections}\label{sec:varpi}
Another crucial difference between the non-Abelian and the Abelian case, concerns the radiative/Coulombic decomposition.
Recall that a crucial ingredient of that decomposition is the duality in $\Omega$ between the ``pure gauge part'' of $A$ and the  radiative part of $E$---which \emph{coincided with} the part of $E$ that does not partake in the Gauss constraint:
\be
\int  E^i_\rad  \dd \pp_i \varsigma \equiv 0 \qquad\leftrightarrow\qquad \pp_i E^i_\rad \equiv 0 \qquad \text{(Maxwell)}.
\ee
However, in the non-Abelian case the pure gauge part of $A$ is $\D\varsigma$ and therefore, since $[\dd , \D] \neq 0$, 
\be
\int E^i_\rad  \dd \D_i \varsigma \equiv 0 \qquad\not\leftrightarrow\qquad \D_i E^i_\rad \equiv 0.
\ee

This hurdle requires more than a simple technical tweak: it points towards a new geometric structure, that we were not forced to tap into in the Abelian case. This is the \emph{functional connection} 
\be
\varpi \in \Omega^1(\A, \Lie(\G)).
\ee

In order to sketch how this connection form comes about, let us start by illustrating how it helps generalizing the radiative/Coulombic decomposition in the non-Abelian case. 
the first crucial feature of this generalization is that it concerns $\dd A$---rather than $A$ itself (cf. \eqref{eq:dressing}):
\be
\dd A = \slashed\dd A^\rad + \D \varpi
\label{eq:SdW-A}
\ee
where (here $\D^2$ denotes the field-\emph{dependent} gauge-covariant Laplacian):\footnote{For this construction to be globally meaningful, in the non-Abelian case one has to remove from $\A$ reducible configurations, i.e. configurations of $\A$ such that admit a ``Killing symmetry''---that is a $\chi\in\Omega^0(R,\mathfrak g)$ such that $\D\chi = 0$. Indeed, at reducible configurations the EBVP defining $\varpi$ has a non-empty kernel. See Footnote \ref{fnt:killing}.\label{fnt:killing2}}
\be
\begin{cases}
\D^i \slashed\dd A_i^\rad = 0 & \text{in }R\\
s^i \slashed\dd A_i^\rad = 0 & \text{at }\pp R\\
\end{cases}
\qquad\text{and}\qquad
\begin{cases}
\D^2 \varpi = \D^i \dd A_i & \text{in }R\\
s^i\D_i \varpi = s^i \dd A_i & \text{at }\pp R\\
\end{cases}
\label{eq:SdW}
\ee
Here, the symbol $\slashed\dd A^\rad$ stands for a 1-form that needs \emph{not} be exact.\footnote{Due to its geometrical origin, it is denoted $\dd_H A$ in \cite{GomesRiello2016,GomesRiello2018,GomesHopfRiello,GomesRiello-quasilocal,AldoNew} . See \cite[Sec.2]{GomesRiello-quasilocal}.} 
In other words, in the non-Abelian case there is no decomposition of $A$ whose differential leads to (\ref{eq:SdW-A}-\ref{eq:SdW}).
Of course, things are different in the Abelian case: comparison of \eqref{eq:SdW} with \eqref{eq:dressing}, shows that in Maxwell theory the SdW/Coulomb connection is exact $\varpi_\text{Max} = \dd \varsigma$. (Indeed, one can show that a connection form $\varpi$ is exact if and only if it is a flat functional connection of an Abelian gauge theory \cite[Sec.2]{GomesRiello-quasilocal}; more on curvature, shortly).

The crucial properties of the 1-form $\varpi$ defined by the previous equations are its covariance and projection properties:\footnote{The covariance equation presented here differs from those of \cite{GomesRiello2016,GomesRiello2018,GomesHopfRiello,GomesRiello-quasilocal,AldoNew} simply because in this article we only consider gauge transformations $\xi$'s such that $\dd \xi = 0$. However, as discussed in those articles, the two definitions are fully equivalent.}
\be
\begin{cases}
\bb L_{\xi^\#} \varpi = [\varpi, \xi] & \text{(covariance)}\\
\bb i_{\xi^\#} \varpi = \xi & \text{(projection)}
\end{cases}
\label{eq:varpiprop}
\ee
Indeed, these two properties alone \emph{define} the notion of \emph{functional} (\emph{Ehresmann}) \emph{connection} 1-form over the space of gauge potentials $\A$---here $\A\to\A/\G$ can be thought as an infinite dimensional fibre bundle,\footnote{For a review of the limitation of this interpretation due to the presence of reducible configurations, and references to the relevant literature, we refer to \cite[Sec.4]{GomesRiello-quasilocal}.}  with orbits $\mathcal O_A = \{ g^{-1}Ag + g^{-1}\d g, g\in \G\}$. 
In this framework, the second of the equations \eqref{eq:SdW} only defines one particular choice of connection form. This choice,  named after Singer\footnote{In its boundary-less version, the SdW connection form was first defined by Singer in \cite{Singer:1978dk}. See \cite{GomesHopfRiello} for a more complete bibliography on the subject.} and DeWitt (SdW) in \cite{GomesHopfRiello}, generalizes Coulomb gauge to the non-Abelian case. As we will see in a moment, however, it does \emph{not} define an actual gauge fixing.

From the defining properties of $\varpi$ \eqref{eq:varpiprop}, it follows that $\slashed\dd A^\rad$ is necessarily horizontal and covariant:\footnote{These equation hold unaltered even for field-dependent gauge-transformations.}
\be
\begin{cases}
\bb L_{\xi^\#} \slashed\dd A^\rad = [\slashed\dd A^\rad, \xi] & \text{(covariance)}\\
\bb i_{\xi^\#} \slashed\dd A^\rad = 0 & \text{(horizontality)}
\end{cases}
\ee
Which means that any gauge-invariant differential form built out of $\slashed\dd A^\rad$ is basic, i.e. projectable onto the reduced space $\A/\G$. \emph{This is why functional connection forms play a central in the theory of symplectic reduction.}

A central geometric feature of any connection form is its curvature 2-form. 
 The general formula for the functional curvature of $\varpi$ is
\be
\bb F = \dd \varpi + \tfrac12 [\varpi\stackrel{\curlywedge}{,}\varpi].
\ee
For the SdW connection \eqref{eq:SdW}, $\bb F\neq0$ and can be computed from 
\be
\begin{cases}
\D^2 \bb F = [ \slashed\dd A^\rad_i\stackrel{\curlywedge}{,}\slashed\dd A^{\rad i}]  & \text{in }R\\
s^i\D_i \bb F = 0 & \text{at }\pp R\\
\end{cases}
\label{eq:SdW-FF}
\ee

\begin{figure}
\begin{center}
\includegraphics[width=.9\textwidth]{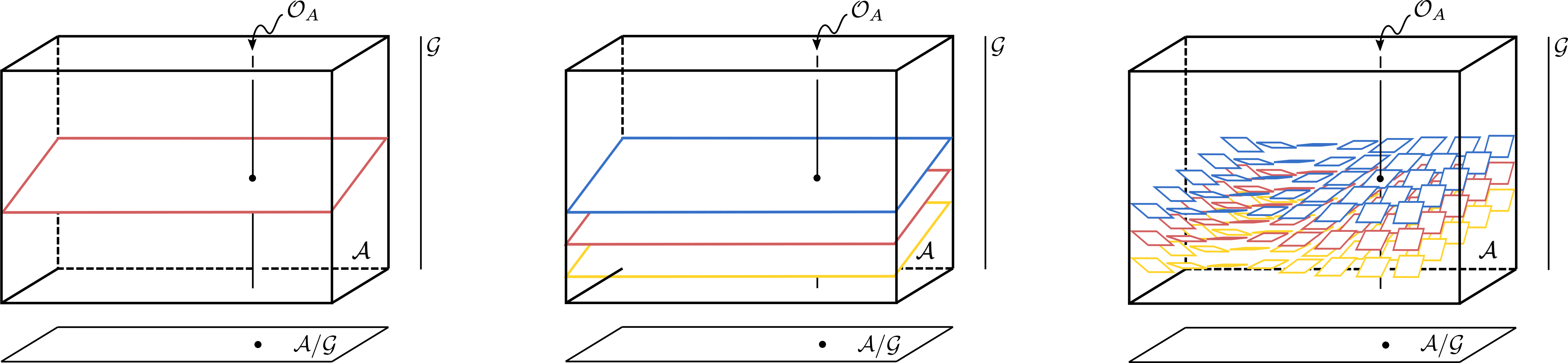}
\caption{
\emph{On the left}: A gauge fixing is a global section of $\A \to \A/\G$. 
\emph{In the center}: A flat functional connection ($\bb F=0$) corresponds to an equivariant family of global sections of $\A \to \A/\G$.
\emph{On the right}: A curved functional connection ($\bb F \neq 0$) corresponds to a non-integrable equivariant ``horizontal'' distribution  in $\A \to \A/\G$; it provides ``infinitesimal gauge fixings'' around each point in field-space which do not extend beyond the tangent-plane approximation. In the presence of a Gribov problem, i.e. whenever global sections of $\A\to\A/\G$ simply do not exist, only connections with curvatures can be globally defined over $\A$.
} 
\label{fig:anholo}
\end{center}
\end{figure}

As standard in the theory of principal fibre bundles, the curvature $\bb F$  measures the (Frobenius) integrability of the so-called ``horizontal'' distribution $H = \mathrm{ker}(\varpi) \subset\T\A$.
Thanks to \eqref{eq:varpiprop}, the distribution $H$ is equivariant under the action of $\G$ and locally defines a complement to the directions tangent to the gauge orbits. 
In this sense, one can loosely say that a connection form defines at every $A$ in $\A$ an ``infinitesimal'' gauge fixing.
More precisely: flat connections correspond to equivariant families of \emph{global} sections $\A/\G \to \A$. E.g. this is the case for the Abelian SdW $\varpi_\text{Max}=\dd \varsigma$, whose section through $A=0$ corresponds to Coulomb gauge fixing of $\A_\text{Max}$. Connections with curvature are \emph{more general} objects than gauge fixings that \emph{can} be defined (like the SdW connection) even when global sections of $\A\to\A/\G$ do not exist, that tis even in the presence of a Gribov-Singer problem \cite{Gribov:1977wm,Singer:1978dk,Singer:1981xw,Zwanziger:1989mf}---see Figure \ref{fig:anholo}.

In the presence of matter fields,  flat connections allow one to introduce dressed matter fields in complete analogy with the Dirac-dressed electron of Section \ref{sec:dressing}. Conversely, to curved connections one can only associate horizontal equivariant differentials $\slashed\dd \hat\psi$, in analogy with $\slashed\dd A^\rad$ above. 
If one allows for non-local objects \emph{in field space}, the SdW connection can be used to introduce ``dressed fields'' defined through certain field-space Wilson lines of $\varpi$ \cite[Sec.5]{GomesRiello-quasilocal}. As discussed in \cite[Sec.9]{GomesHopfRiello}, these field-space Wilson lines, and the ``dressed fields'' they produce,\footnote{That these ``dressing'' cannot in general be global over field space follows from the Gribov problem \cite{Gribov:1977wm, Singer:1978dk, Singer:1981xw, Zwanziger:1989mf}. The same limitation must hold in the Vilkovisky--DeWitt formalism as well. See  \cite[Sec.9]{GomesHopfRiello}.} are strictly related to ideas developed by Vilkovisky and DeWitt in the context of the so-called ``geometric effective action'' for gauge theories \cite{vilkovisky1984gospel,Vilkovisky:1984st,DeWitt_Book} and to DeWitt's ghost-free formulation of gauge theories \cite{DeWitt:1995cx}.\footnote{We also refer to \cite{Pawlowski:2003sk,Branchina:2003ek}, which discuss the consequences for the geometric effective action of the residual gauge ambiguity present in the Vilkovisky--DeWitt formalism. Expressed in terms of the functional-connection, this ambiguity is related to the choice of reference configuration from which the dressing-defining field-space Wilson lines are chosen to depart. As such, this ambiguity is (implicitly) present in the Dirac dressing equation \ref{eq:dressedpsi} as well \cite[Sec.9]{GomesHopfRiello}.}

\paragraph{The Gauss constraint}
As already mentioned in the context of Maxwell theory, to any radiative/pure-gauge decomposition of $\dd A$ there corresponds a dual radiative/Coulombic decomposition of $E$. The SdW decompositon of $E$ reads
\be
E^i = E^i_\rad + E^i_\Coul,
\qquad\text{with}\qquad
\begin{cases}
\D_i E^i_\rad = 0 & \text{in }R\\
s_i E^i_\rad = 0 & \text{at }\pp R
\end{cases}
\ee
Different choice of functional connections do not modify the functional properties of $E^\rad$ (see Rermark \ref{Rmk:fcoord}), but lead to different functional forms of the Coulombic part of the electric field $E_\Coul = E - E_\rad$ through the duality condition $\int \tr(E^i_\Coul \curlywedge \slashed\dd A^\rad_i) \equiv 0$. Any two such instantiations of $E_\Coul$  necessarily differ by a radiative electric field. 
One can show \cite{AldoNew} that for all choices of $\varpi$ the Gauss constraint---equipped with the appropriate boundary condition\footnote{Again, this is the case only if reducible configurations are removed from $\A$. Cf. Footnote \ref{fnt:killing2}. }
\be
\begin{cases}
\D_iE^i_\Coul = 0 & \text{in }R\\
s_i E^i = f & \text{at }\pp R
\end{cases}
\label{eq:nonAbGauss}
\ee
completely fixes $E_\Coul$ throughout $R$ as a function of $f$.

Going back to the SdW choice of connection, in complete analogy with the Abelian case, it is easy to check that the SdW connection leads to a ``self-dual'' decomposition of $\dd A$ and $E$, where the two decompose along the same functional bases.
That is, for the SdW choice of $\varpi$, 
\be
E^i_\Coul = \D^i \varphi.
\ee
and the Gauss constraint turns into the same EBVP that fixes the SdW connection and its curvature:
\be
\begin{cases}
\D^2\varphi = 0 & \text{in }R\\
s_i \D^i\varphi = f & \text{at }\pp R
\end{cases}
\ee

Summarizing, \emph{for any given choice of $\varpi$}, the flux $f$ completely fixes $E^i_\Coul$ throughout $R$. 
Therefore, on-shell, $(E^i_\rad , f)$ constitute a complete set of coordinates on the space of electric fields.
Different choice of $\varpi$ correspond to different choices of ``axes'' along which these coordinates are defined.

\paragraph{Flux rotations}
As already observed, the flux $f$ is not completely fixed in a superselection sector $\Phi_o^{[f_o]}$. 
One can therefore ask what kind of changes corresponds to changes, or ``rotations,'' of $f$ within $[f_o]$ \textit{at fixed values of $E_\rad$ and $A$}:
\be
\delta_\zeta f = [f ,\zeta],
\qquad 
\delta_\zeta E^i_\rad = 0 = \delta_\zeta A.
\label{eq:fluxrot}
\ee
labelled by a $\zeta$ valued in $C(\pp R, \frak g)$.
Note that the definition of flux rotations \emph{depends} on the choice of a radiative/Coulombic split, i.e. on a choice of $\varpi$.

Since, given a choice of $\varpi$, $f$ parametrizes a component of the electric field throughout $R$, flux rotations actually affect the value of the electric field \emph{throughout} $R$.
However, as already emphasized, the precise way the electric field changes under a given change of $f$ depends on the choice of $\varpi$. 
For this reason, whether or not flux rotations are Hamiltonian depends on the choice of $\varpi$ through which flux rotations are defined. In particular, flux rotations can be Hamiltonian symmetries of $(\Phi //_{\hspace{-2pt}[f_o]\hspace{2pt}}, \Omega^{[f_o]}_r)$ \emph{if and only if} $f$ refers to a choice of a \emph{flat} $\varpi$. 
In this case, the Hamiltonian generation of the flux rotation \eqref{eq:fluxrot} is
\be
Q[\zeta] = \oint \tr(\zeta f).
\ee
(Important subtleties lurk behind this simple expression, which are related to the curvature of $\varpi$; see \cite[Sec.4.12]{AldoNew}.)

Notice that, although $f$ changes as in a gauge transformation, flux rotations are not gauge transformations. Indeed, they do \emph{not} affect either $E^i_\rad$ nor $A_\rad$. As a consequence, even the Coulombic electric field itself, i.e. $E^i_\Coul=E^i_\Coul(f)$ as computed from \eqref{eq:nonAbGauss}, although affected, does \emph{not} undergo an adjoint transformation. This means in particular that flux rotations do affect the total energy content of the region $R$ and are therefore \emph{not} expected to be dynamical symmetries. 

Therefore, despite appearances, flux rotations are very different from the ``boundary symmetries'' described in e.g. \cite{DonnellyFreidel} (and many follow-up publications).

Summarizing, flux rotations are a new set of \emph{physical} transformations of the field configuration \emph{throughout} $R$ which manifests itself in the non-Abelian theory. Their precise definition in the bulk depends on a choice of $\varpi$. In fact, they can be Hamiltonian symmetries \emph{if only if} they refer to a \emph{flat} functional connection $\varpi$, in which case their Hamiltonian generator---in the corresponding functional coordinates---is the smeared flux $Q[\zeta]$. 
Since flux rotations affect the total Hamiltonian of the theory, they are \emph{not} expected to be dynamical symmetries.

\paragraph{Geometric ghosts and BRST}
We conclude the article with a brief comment about the geometrization of BRST.
``Geometrization of BRST'' refers to a program of finding a geometric interpretation for the otherwise algebraic machinery which is BRST. 
This program was initiated by Thierry-Mieg and collaborators in the late 1970s \cite{Thierry-MiegJMP, thierry1985classical,BAULIEU1982477} (see also the recent \cite{CiambelliLeigh:BRST}, and references therein). 

The very basic idea behind their pioneering work was to interpret the BRST operator as a vertical\footnote{Loosely speaking, here ``vertical'' means ``along the fibres of the bundle.''} differential in the \emph{finite} dimensional bundle $P\to \Sigma$ and the ghost as a vertical Maurer-Cartan form along its fibres $F\cong G$: this way the BRST transformation of the ghost, $\mathsf{s} c = \tfrac12[c,c]$ is nothing else than Cartan's second structural equation. But in this simple form their idea was prone to drawbacks \cite{LeinaasOlaussen:BRST:1981}, and was therefore subsequently modified. 

One of these modifications called for the replacement of the finite dimensional bundle $P\to \Sigma$, on which $A$ corresponds to an Ehresmann connection, to the \emph{infinite dimensional} bundle $\A \to \A/\G$, in which $A$ is a point \cite{LeinaasOlaussen:BRST:1981, Bonora1983}.
Then, in this infinite dimensional transposition of the Thierry-Mieg construction, the BRST operator would become the vertical functional differential $\dd_V$ and the ghost the Maurer-Cartan form $\vartheta$ along the \emph{infinite} dimensional fibres $\cal F \cong \G$. 

One of the advantages of this viewpoint was to focus the attention on $\dd_V$ over $\G$, whose cohomology was first shown by Bonora and Cotta-Ramusino to carry crucial information about the gauge theory, in particular about the chiral anomaly \cite{Bonora1983}.\footnote{See e.g. \cite[Sect.2]{Barnich:2000zw} for a historical overview of the cohomological computation of gauge anomalies---and much more.}
Furthermore, as discussed by Strocchi \cite[Ch.3]{Strocchi_2019}, the link between the topology of the infinite dimensional group $\G$ and chiral symmetry breaking can be drawn in a rigorous algebraic setting by studying the properties of the (centre of the) algebra of gauge-invariant observables under chiral transformations.\footnote{In \cite[Ch.3]{Strocchi_2019}, instanton configurations and \emph{their} topology play no role.}

In \cite{GomesRiello2016} with Gomes, we put forward a slight generalization of the proposal of Bonora and Cotta-Ramusino, in which $\varpi$ plays the role of the ghost. 
Since the pullback of any $\varpi$ to a fibre $\cal F \cong \G$, \emph{is} the Maurer-Cartan form $\vartheta$, as far as the action of the geometric BRST operator $\dd_V$ is concerned, there is no difference between the new and the old proposals.
However, the new proposal comes with two advantages: \i  whereas $\vartheta$ is only defined on each \emph{individual} fibre $\cal F$, $\varpi$ is defined globally across $\A$; and \ii the freedom of choosing $\varpi$---i.e. a gauge fixing (or its non-integrable generalization)---seems to nicely relate to the different choices of Faddeev--Popov operator. To the best of my knowledge, the latter relation was so far left unexplained by other proposals to geometrize BRST.

\newpage

\footnotesize 
 \bibliographystyle{bibstyle_aldo}

\end{document}